\documentclass[aps,prx,twocolumn,superscriptaddress,english,floatfix]{revtex4-2}
\usepackage{graphicx}
\usepackage{float}
\usepackage{physics}
\usepackage{cancel}
\usepackage{overpic}
\usepackage{enumerate}
\usepackage{dsfont}

\usepackage[colorlinks,citecolor=blue,linkcolor=blue,urlcolor=blue]{hyperref}
\usepackage{textcomp}
\usepackage{amsmath}
\usepackage{amssymb}
\usepackage{soul}
\usepackage[normalem]{ulem}
\usepackage{lipsum}
\usepackage{comment}
\usepackage{mathrsfs}

\usepackage[english]{babel}
\usepackage{bm}
\usepackage{orcidlink}

\graphicspath{{./}{./figs/}}

\let\originalleft\left
\let\originalright\right
\renewcommand{\left}{\mathopen{}\mathclose\bgroup\originalleft}
\renewcommand{\right}{\aftergroup\egroup\originalright}

\newcommand{\vect}[1]{\boldsymbol{#1}}
\renewcommand{\vec}[1]{\vect{#1}}

\makeatletter
\newcommand*\bigcdot{{\color{gray}\mathpalette\bigcdot@{1.}}}
\newcommand*\bigcdot@[2]{\mathbin{\vcenter{\hbox{\scalebox{#2}{$\m@th#1\bullet$}}}}}
\makeatother

\def\rme{{\rm {e}}}
\def\rmi{{\rm {i}}}
\renewcommand{\d}{{\rm {d}}}
\def\dt{{\rm {d}}t}

\newcommand{\idhat}{\hat{\mathds{1}}}

\newcommand{\alphatil}{{\Tilde{\alpha}}}

\newcommand{\betatil}{{\Tilde{\beta}}}

\newcommand{\delhat}{\hat{\delta}}
\newcommand{\ddelhat}{\hat{\delta}^\dagger}

\newcommand{\jcal}{\mathcal{J}}
\newcommand{\Jcal}{\jcal}

\newcommand{\rhohat}{\hat{\rho}}

\newcommand{\sigmahat}{\hat{\sigma}}

\newcommand{\sinc}{\mathrm{sinc}}

\newcommand{\schr}{Schr\"{o}dinger}

\newcommand{\hc}{\mathrm{H.c.}}

\newcommand{\Ahat}{\hat{A}}

\newcommand{\bbb}{\hat{b}}
\newcommand{\dbbb}{\hat{b}^\dagger}

\newcommand{\Bhat}{\hat{B}}

\newcommand{\hcal}{\mathcal{H}}

\newcommand{\Hhat}{\hat{H}}

\newcommand{\lcal}{\mathcal{L}}
\newcommand{\Lhat}{\hat{L}}
\newcommand{\dLhat}{\Lhat^\dagger}

\newcommand{\Lcal}{\mathcal{L}}

\newcommand{\mvec}{{\vec{m}}}

\newcommand{\ncal}{\mathcal{N}}

\newcommand{\nvec}{{\vec{n}}}
\newcommand{\nhat}{{\hat{n}}}

\newcommand{\Ohat}{\hat{O}}

\newcommand{\Qhat}{\hat{Q}}

\newcommand{\Rhat}{\hat{R}}

\newcommand{\rvec}{\vec{r}}

\newcommand{\rf}{{\mathrm{RF}}}

\newcommand{\shat}{\hat{s}}

\newcommand{\xtil}{{\Tilde{x}}}

\newcommand{\ytil}{{\Tilde{y}}}

\newcommand{\ztil}{{\Tilde{z}}}

\newcommand{\bea}{\begin{equation}\begin{aligned}}
		\newcommand{\eea}{\end{aligned}\end{equation}}
\newcommand{\be}{\begin{equation}}
	\newcommand{\ee}{\end{equation}}

\begin{document}

\title{Generalized stochastic spin-wave theory for open quantum spin systems}

\author{Zejian Li~\orcidlink{0000-0002-5652-7034}}
\affiliation{The Abdus Salam International Center for Theoretical Physics, Strada Costiera 11, 34151 Trieste, Italy}
\affiliation{Universit\'{e} Paris Cit\'e, CNRS, Mat\'{e}riaux et Ph\'{e}nom\`{e}nes Quantiques, 75013 Paris, France}
\author{Anna Delmonte~\orcidlink{0009-0008-9371-6855}}
\affiliation{SISSA, Via Bonomea 265, I-34136 Trieste, Italy}
\affiliation{JEIP, UAR 3573 CNRS, Coll\`{e}ge de France, PSL Research University,
11 Place Marcelin Berthelot, F-75321 Paris, France}

\author{Rosario Fazio~\orcidlink{0000-0002-7793-179X}}
\affiliation{The Abdus Salam International Center for Theoretical Physics, Strada Costiera 11, 34151 Trieste, Italy}
\affiliation{Dipartimento di Fisica ``E. Pancini", Universit\`a di Napoli ``Federico II'', Monte S. Angelo, I-80126 Napoli, Italy}

\begin{abstract}
    We propose a semiclassical framework for solving open quantum dynamics in driven-dissipative spin systems. Our method consists of generalized spin-wave approximations tailored to describing quantum trajectories unravelled from the master equation, and generically applies to regimes beyond the reach of conventional spin-wave theories, including short-range interactions and local quantum jumps, enabling the efficient simulation of large-scale interacting spins. We illustrate the versatility of our framework by studying a variable-range driven-dissipative Ising model on a 2D lattice. When the dissipation acts along the drive axis, we find a continuous phase transition breaking the $\mathbb{Z}_2$ symmetry, and demonstrate that the interaction range, when tuned from fully-connected to nearest-neighbour, profoundly alters the universality class of the criticality. With the dissipation along the interaction axis, we show the emergence of a first-order transition. Demonstrated with both state-diffusion and quantum-jump types of trajectory dynamics, our framework provides a powerful toolbox for the efficient semiclassical description of non-equilibrium dynamics and many-body phases in spin systems.
\end{abstract}

\maketitle



\section{Introduction}

A central frontier in modern physics is the study of open quantum many-body systems~\cite{fazioManybodyOpenQuantum2025,breuer2002theory}, where an exciting variety of out-of-equilibrium phases and critical phenomena occur due to the rich interplay between unitary quantum dynamics and the dissipative environment. 
Along with their importance for quantum technologies~\cite{Preskill2018quantumcomputingin, Wiseman}, the recent decades have witnessed impressive development in experimental platforms~\cite{Greiner, Gross2017, Monroe_2021, blaisCircuitQuantumElectrodynamics2021}, where both the interaction and dissipation can be engineered~\cite{blaisQuantumInformationProcessing2020, Underwood, Safavi2018,Cole2021,leghtas2015confining}. They offer a versatile playground for exploring nonequilibrium quantum dynamics~\cite{Fitzpatriv,Pieczarka,Castellanos, Zejian, Mei}, and, in turn, sparked intense theoretical activities~\cite{Sieberer_2016, Eisert_2015, Camacho_2022, Vasseur_2016, Bernard_2016}.

The investigation of open quantum many-body systems is, in general, a formidable challenge. Besides the rare cases of analytically solvable models, see e.g.~\cite{foss-feigSolvableFamilyDrivenDissipative2017,paz2021,bartoloExactSteadyState2016}, one generally needs to resort to numerical simulations or approximate approaches. Approximation methods in particular, exploit prior knowledge on the physical system and reduce the dynamics to a much smaller effective manifold via clever design of the state representation.

In systems with long-range magnetic ordering, the \textit{spin-wave} approximation proved to be useful in many different contexts~\cite{herringTheorySpinWaves1951,kuboSpinWaveTheoryVariational1953,vankranendonkSpinWaves1958}, based on a semiclassical representation of spins via the Holstein-Primakoff transformation~\cite{blochZurTheorieAustauschproblems1932,holsteinFieldDependenceIntrinsic1940}. To the lowest order, this can be regarded as approximating the quantum fluctuations on top of a classical state as a set of harmonic oscillators. In equilibrium settings, the spin-wave theory has been highly successful in capturing the magnetic properties of a wide range of materials, showing good accordance with experimental data~\cite{vankranendonkSpinWaves1958}. It has also been extended to the time-dependent regime for both unitary and dissipative systems~\cite{ruckriegelTimedependentSpinwaveTheory2012a,leroseImpactNonequilibriumFluctuations2019,seetharamDynamicalScalingCorrelations2022} to describe non-equilibrium dynamics~\cite{defenuOutofequilibriumDynamicsQuantum2024}.  

More recently, the spin-wave framework has been generalized by us to describe quantum trajectories in driven-dissipative long-range systems, which was formulated in the context of heterodyne monitoring of a collective decay channel~\cite{liMonitoredLongrangeInteracting2025a} and shown to provide a sound approximation of the monitored dynamics. A distinct advantage comes from the stochastic nature of the approach: while the physical state \textit{along the quantum trajectory} is approximated by a Gaussian ansatz~\cite{verstraelen2018gaussian, delmonteMeasurementinducedPhaseTransition2026}, which captures quantum correlations on the quadratic level, the trajectory-averaged state, consisting of a mixture of Gaussians, provides a much more accurate representation of the Lindblad dynamics compared to deterministic semiclassical methods targeting directly the density matrix~\cite{seetharamDynamicalScalingCorrelations2022}, allowing us to capture highly non-Gaussian mixed states that are by construction lost in the latter.  

In this work, we reformulate and further generalize the framework of spin-wave quantum trajectories (SWQT) introduced in~\cite{liMonitoredLongrangeInteracting2025a}, making it a generic semiclassical approach to capturing the driven-dissipative spin dynamics, applicable to regimes including local quantum jumps and short-range interactions, where the conventional spin-wave approaches typically break down. This is achieved via generalized spin-wave approximations along quantum trajectories, featuring two essential new ingredients:
\begin{itemize}
    \item we bosonize each individual spin in its \textit{local} comoving frame with a higher-order Holstein-Primakoff expansion, and the bosonic state is approximated with a Gaussian ansatz;
    \item we introduce a non-singular parametrization of the comoving frames via \textit{quaternions}, resolving the common pathology of spherical coordinate singularities affecting all time-dependent spin-wave theories~\cite{ruckriegelTimedependentSpinwaveTheory2012a,leroseImpactNonequilibriumFluctuations2019,seetharamDynamicalScalingCorrelations2022}.
\end{itemize}
  The resulting spin-wave dynamics can be elegantly cast in a stochastic differential equation, enabling efficient simulation of the dynamics, for both \textit{state-diffusion} and \textit{quantum-jump} types of monitoring. The method is carefully benchmarked, and we show that it is exact in the non-interacting limit, consistent with the semiclassical assumptions. To demonstrate the wide applicability of the framework, we study a two-dimensional (2D) transverse-field Ising model subjected to local dissipation, with a power-law interaction profile allowing the interaction range to be tuned from fully-connected to nearest-neighbor. When the dissipation acts along the drive axis, the steady state admits a $\mathbb{Z}_2$ symmetry-breaking phase transition, and we show that the interaction range profoundly alters the dissipative criticality in this system, with the critical exponent moving from the mean-field value to that of the 2D Ising universality class. We also consider a variant of the model with the dissipation acting along the interaction direction, and show the emergence of a first-order phase transition.

The rest of the paper is structured as follows. In Sec.~\ref{sec:long-range-local-diss-model}, we define the power-law spin model under consideration and its quantum-trajectory dynamics. Sec.~\ref{sec:swqt-new} contains the core theory of our work, where we construct the framework of the generalized stochastic spin-wave theory and derive the semiclassical representation of the quantum trajectories. The method is then applied to the power-law spin system in Sec.~\ref{sec:results}, where we discuss results on the driven-dissipative phases. Finally, we conclude in Sec.~\ref{sec:conclusions} with some perspectives.

\section{Quantum trajectories in dissipative spin lattices}\label{sec:long-range-local-diss-model}

We consider a lattice of interacting spins, whose dissipative dynamics can be cast in the form of a Lindblad master equation~\cite{breuer2002theory},
\bea\label{eq:lindblad-me}
    \dfrac{\d\rhohat}{\d t} &= -\rmi[\Hhat,\rhohat] + \sum_i\left( \Lhat_i\rhohat\dLhat_i - \dfrac{1}{2}\left\{\dLhat_i\Lhat_i,\rhohat\right\} \right)\\
    &= \lcal\rhohat\,,
\eea
where $\rhohat$ is the density matrix representing the system's mixed state, $\Hhat$ is the Hamiltonian, $\Lhat_i$ is the Lindblad jump operator and $\lcal$ denotes the Liouvillian superoperator generating the dissipative evolution, which we now specify.
While the formalism of our spin-wave framework will not depend on the particular form of the model, we expect the theory to be best applicable to systems with ``semiclassical" behavior, such as macroscopic collective ordering, as is the case in many long-range interacting systems. For this motivation, we will consider in this work a driven-dissipative spin model with a power-law interaction profile, as sketched in Fig.~\ref{fig:spins} (a). The model is defined on a two-dimensional periodic lattice of $N=L\times L$ spins at positions $\rvec_i=(r^{x}_i, r^{y}_i)$ with $r_i^\mu=1,\ldots,L$.

We will consider two variants of the model. In the first model, the Hamiltonian is defined as follows,
\bea\label{eq:ham-zxx}
\mathrm{Model~ I:~}    \Hhat^{z-xx} = h\sum_i\sigmahat^z_i - \sum_{i\neq j} J^{(\alpha)}_{ij}\sigmahat^x_i\sigmahat^x_j\,,
\eea
where $\sigmahat^\mu_i$ are Pauli matrices for spin-$1/2$, $h$ represents an external magnetic field, $J^{(\alpha)}_{ij}$ is the interaction strength between sites $i$ and $j$, where we assume a power-law dependence on the distance,
\bea
    J^{(\alpha)}_{ij} = \dfrac{J}{\ncal_\alpha}\dfrac{1}{\lVert \rvec_i-\rvec_j \rVert^\alpha}\,,
\eea
with the exponent $\alpha$ controlling the interaction range~\footnote{Note that with periodic boundary conditions, the distance is defined as
$\lVert \vec{r}_i - \vec{r}_j\rVert\equiv  \sqrt{\sum_{\mu=x,y}\min(|r_i^{\mu}-r_j^{\mu}|,L-|r_i^{\mu}-r_j^{\mu}|)^2}$.}, and the Kac normalization $\ncal_\alpha=\frac{1}{N}\sum_{i\neq j}\lVert\rvec_i-\rvec_j\rVert^{-\alpha}$ ensures a well-defined thermodynamic limit~\cite{kacVanWaalsTheory1963}. In particular, $\alpha= 0$ implies an infinite range, and the limit of $\alpha\to\infty$ corresponds to nearest-neighbor interactions.
 We model the dissipation with local spin decay, defined by the Lindblad operator \bea
 \Lhat_i=\sqrt{\gamma}\sigmahat^-_i\,,
 \eea
 with $\sigmahat^\pm\equiv\sigmahat^x_i\pm\rmi\sigmahat^y_i$, which acts along the drive axis ($z$). The Liouvillean therefore admits a $\mathbb{Z}_2$ symmetry due to its invariance under the transformation $\sigmahat^{x,y}_i\to-\sigmahat^{x,y}_i$. In a mean-field theory (see Appendix.~\ref{app:mf}), this symmetry can be spontaneously broken,
resulting in an ordered phase with nonzero magnetization along $x$ and $y$ [see, for example, the mean-field phase boundary in Fig.~\ref{fig:x2_ent_heatmap_6} (a)]. 

The second model we consider has a modified Hamiltonian, with the interaction and driving axes interchanged,
\bea\label{eq:ham-xzz}
    \mathrm{Model~ II:~} \Hhat^{x-zz} = h\sum_i\sigmahat^x_i + \sum_{i\neq j} J^{(\alpha)}_{ij}\sigmahat^z_i\sigmahat^z_j\,,
\eea
and the dissipation operator $\Lhat_i=\sqrt{\gamma}\sigmahat_i^-$ remains the same, which now acts along the interaction axis. This setting no longer has the $\mathbb{Z}_2$ symmetry in the Liouvillian, and the mean-field theory predicts a bistable region~\cite{jinPhaseDiagramDissipative2018} for large enough interaction strength $J$ [see, for example, the mean-field solution in Fig.~\ref{fig:xzz-qj-v3}].

As a key ingredient of our framework, we \textit{unravel} the master equation~\eqref{eq:lindblad-me} into \textit{quantum trajectories}, which allows approximating highly non-Gaussian mixed states via ensemble averages of (Gaussian) single trajectories. This unraveling process corresponds to an alternative description of open quantum dynamics by modeling the dissipative environment as an observer that continuously monitors the system with weak measurements, resulting in stochastic dynamics of the system state. In this work, we will consider two typical unravelings usually considered in the quantum optics community, the \textit{heterodyne} unraveling and the \textit{quantum-jump} unraveling, which correspond to different measurement protocols applied to the output current from the open system, while their ensemble-averaged mixed state follows the same Lindblad dynamics.

\subsection{Heterodyne}

The first unraveling we will consider in this work describes the measurement process of heterodyne detection, which is given by the following stochastic equation,
\bea\label{eq:het-drho}
    \d\rhohat = \d t\lcal\rhohat + \d\hcal\rhohat\,,
\eea
where the second term encodes the stochastic noise arising from the monitoring,
\bea
	\d\hcal\rhohat&\equiv \sum_i\hcal[\d Z^*_i\Lhat_i]\rhohat\\&=\sum_i \left[\d Z_i^*(\Lhat_i-\langle\Lhat_i\rangle)\rhohat + \hc\right]\,,
\eea
with $\d Z_i$ the complex Wiener process defined via its second moment,
\bea\label{eq:dZ-het}
	\d Z_i\d Z_j = 0\,,\quad
	\d Z_i^*\d Z_j = \delta_{ij}\d t\,,
\eea
and $\langle\bullet\rangle\equiv\Tr[\rhohat\bullet]$ is the \textit{single-trajectory} expectation value, whose dependence on the current state $\rhohat$ makes the dynamics nonlinear, accounting for the measurement back action.
The heterodyne dynamics is continuous in time and describes a quantum state diffusion process.

\subsection{Quantum jump}
We will also consider the quantum-jump unraveling, which models the experimental process of counting photons emitted from the system, whose dynamics can be cast in the stochastic master equation as follows,
\bea\label{eq:qj-rho}
    \d\rhohat &= \d t\left\{ -\rmi[\Hhat,\rhohat] - \dfrac{1}{2}\sum_i\hcal[\dLhat_i\Lhat_i]\rhohat \right\}\\
    &+ \sum_i \d M_i(t)\Jcal[\Lhat_i]\rhohat\,,
\eea
where we denote
\bea
    \hcal[\Ahat]\rhohat &\equiv \left( \Ahat-\langle\Ahat\rangle \right)\rhohat + \rhohat \left( \Ahat^\dagger - \langle\Ahat^\dagger\rangle \right)\,,\\
    \Jcal[\Lhat]\rhohat&= \dfrac{\Lhat\rhohat\dLhat}{\langle\dLhat\Lhat\rangle}-\rhohat\,,
\eea
and $M_i(t)$ represents the number of jumps occured up to time $t$ corresponding to the jump operator $\Lhat_i$. The point process $\d M_i(t)$ satisfy
\bea\label{eq:dm-jump}
    \d M_i(t) \d M_j(t) &= \d M_i(t) \delta_{ij}\,,\\\quad \mathbb{E}[\d M_i(t)] &= \langle\Lhat_i^\dagger\Lhat_i\rangle(t)\d t\,.
\eea
This implies that each increment $\d M_i(t)$ can only take values in $\{0,1\}$, where $\d M_i=1$ corresponds to the event of a discontinuous jump, whose probability at each infinitesimal time step $\d t$ is $\langle\Lhat_i^\dagger\Lhat_i\rangle\d t$. The contrary case of $\d M_i=0$ then describes a continuous state evolution governed by an effective Hamiltonian
\bea\label{eq:Heff-nh}
    \Hhat_\mathrm{eff} &= \Hhat - \dfrac{\rmi}{2}\sum_i\dLhat_i\Lhat_i\,,
\eea
which is non-Hermitian and governs the no-jump dynamics.
Note that in the stochastic master equations for both unravelings~\eqref{eq:het-drho} and~\eqref{eq:qj-rho}, the density matrix $\rhohat$ represents the state conditioned on the trajectory instead of the trajectory-averaged mixture, and one can verify using the respective noise definitions that~\cite{Wiseman}
\bea
    \d\mathbb{E}[\rhohat] = \dt \lcal\mathbb{E}[\rhohat]\,,
\eea
i.e., the average-state dynamics is indeed governed by the same Lindblad master equation~\eqref{eq:lindblad-me} for both unravelings.

\begin{figure*}
    \centering
    \includegraphics[width=0.9\linewidth]{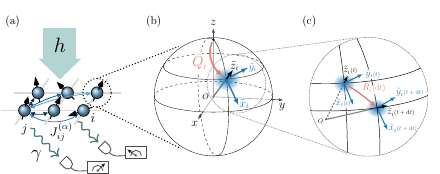}
    \caption{Semiclassical representation of driven-dissipative spin dynamics with the generalized framework of spin-wave quantum trajectories (SWQT). (a) Sketch of a variable-range interacting spin model on a two-dimensional (2D) lattice. The spins are driven with an external field $h$ and interact via a variable-range coupling $J_{ij}^{(\alpha)}$, whose strength decays as a power law of the distance with exponent $\alpha$. The spins are subjected to local decay (at rate $\gamma$), and the dissipative dynamics is unraveled into quantum trajectories via continuous monitoring. (b) In our semiclassical treatment, each spin is bosonized around its classical polarization. Here, $Oxyz$ depicts the lab frame and $O\xtil_i\ytil_i\ztil_i$ is the comoving frame of the $i-$th spin (related to the lab frame via a passive rotation $Q_i$), where the $\ztil_i$ axis aligns with the spin's magnetization vector $\langle\vec{\sigmahat}_i\rangle$. (c) The coupled evolution of the bosonized state and the local frame $Q_i$ along the stochastic trajectory dynamics. At every time step $\d t$, the local frame $Q_i$ is updated with a rotation $R_i(\d t)$ parallel-transporting the comoving frame basis, to preserve the alignment with the spin's orientation without any singularity in the parametrization.}
    \label{fig:spins}
\end{figure*}

\section{Generalized stochastic spin-wave theory along quantum trajectories}\label{sec:swqt-new}

In this section, we formulate the generalized framework of spin-wave quantum trajectories (SWQT), a semiclassical approximation method tailored for describing driven-dissipative spin dynamics, which further extends the stochastic spin-wave theory developed in our previous work~\cite{liMonitoredLongrangeInteracting2025a}. 
In Sec.~\ref{sec:local-sw}, we first bosonize spin operators in \textit{local} comoving frames for each individual spin, which, a priori, admit independent orientations, as graphically represented in Fig.~\ref{fig:spins} (b). This accommodates most generic spin systems with local spin damping and no longer relies on the assumption that the system state always exhibits a well-defined collective polarization along the trajectory dynamics.

Secondly, we present in Sec.~\ref{sec:quat} the quaternion formalism for parameterizing the comoving frames (in which the spins are bosonized),  thereby resolving the common pathology of coordinate singularities shared by all previous time-dependent spin-wave theories. In a conventional spin-wave approach, one typically performs the bosonization of the spin operators in a comoving frame parametrized by angle coordinates of the Bloch sphere $(\theta,\phi)$ with predefined local frame orientations, which inevitably results in coordinate singularities -- a direct consequence of the \textit{hairy ball theorem}~\cite{brouwerUberAbbildungMannigfaltigkeiten1912} which states that one cannot have a nonvanishing continuous tangent vector field (in this case, the local frames are defined by tangent unit vectors) on a 2-sphere. In practice, this hinders the efficient solution of the dynamics if the spin polarization approaches a coordinate singularity. We completely remove this obstacle by introducing a non-singular parametrization of the local frames using quaternions and by parallelly transporting the comoving frame along the dynamics [Fig.~\ref{fig:spins} (c)]. The resulting spin-wave dynamics can then be formulated as a stochastic differential equation, allowing for efficient simulations of the quantum trajectories.

\subsection{Local spin-wave approximations}\label{sec:local-sw}

Our semiclassical approach builds on the assumption that each individual spin in the system admits a well-defined classical magnetization direction (which is, for example, the case of spin coherent states) along the trajectory dynamics~\footnote{Note that this condition can be more easily satisfied along trajectories than on the level of the average mixed state.}. This local spin orientation can be parametrized by the rotation operator $\Qhat_i\equiv\rme^{-\rmi\theta_i\nvec_i\cdot\hat{\vec{s}}_i}$, for some generic rotation angle $\theta_i$ and rotation axis $\vec{n}_i$, and $\vec{\shat}_i=(\shat^x_i,\shat^y_i,\shat^z_i)$ denotes the spin operators (in the lab frame $Oxyz$) on the site $i$, which satisfy the standard $\mathfrak{su}(2)$ algebra $[\shat_i^\mu,\shat^\nu_j]=\rmi\delta_{ij}\varepsilon^{\mu\nu\lambda}\shat^\lambda_i$. For spin $s=1/2$, they are related to the Pauli matrices via $\shat_i^\mu = \sigmahat_i^\mu/2$. This rotation allows us to define a \textit{comoving frame} $O\xtil_i\ytil_i\ztil_i$ [see Fig.~\ref{fig:spins} (b)] aligned with the spin orientation, in which the comoving spin operators are defined via the passive transformation $\shat^\alphatil_i=\Qhat^\dagger_i\shat_i^\alpha\Qhat_i=\sum_\beta Q^i_{\alphatil\beta}\shat_i^\beta$, where we denote the matrix representing the rotation $\Qhat_i$ as $Q^i_{\alphatil\beta}$ when there is no ambiguity. For the moment, we leave the rotation $\hat{Q}_i$ in its most general form, and only assume that it satisfies the alignment condition that $\langle\shat_i^\xtil\rangle=\langle\shat_i^\ytil\rangle=0$, i.e. the comoving $\ztil$ axis be parallel to the local magnetization vector $\langle\vec{\shat}\rangle$.  The comoving spin operators are then bosonized via the Holstein-Primakoff transformation,
\bea\label{eq:bosonization-n}
\shat_i^{\widetilde{+}} = \sqrt{2s}f(\nhat)\bbb_i\,,~\shat_i^{\widetilde{-}}=\sqrt{2s}\dbbb f(\nhat_i)\,,~\shat_i^\ztil=s-\dbbb_i\bbb_i\,,
\eea
where $\shat_i^{\widetilde{\pm}}\equiv \shat_i^\xtil\pm\rmi\shat^\ytil_i$, $\bbb_i$ are bosonic operators satisfying $[\bbb_i,\dbbb_j]=\delta_{ij}$, $\nhat_i\equiv\dbbb_i\bbb_i$, and the bosonic operator function is defined as 
\bea\label{eq:fn}
    f(\nhat) = \sqrt{1-\dfrac{\nhat}{2s}}\,.
\eea
While this mapping is exact, one typically needs to expand the square root as a series and truncate it at a certain order to make practical use of this formalism. The lowest order yields $f_0=1$, which defines the regime of the standard linear spin-wave theory, valid for regimes where $\langle\nhat\rangle/2s\ll1$. In this work, we will consider the next-to-leading-order truncation in a Newton series expansion~\cite{konigNewtonSeriesExpansion2021},
\bea
    f(\nhat)\simeq 1 - \left( 1 - \sqrt{1-\dfrac{1}{2s}} \right)\nhat\,.
\eea
In the case of $s=1/2$, this mapping is exact and represents the hard-core boson limit,
\bea\label{eq:bosonize-sigma}
    \sigmahat_i^{\widetilde{+}}&= 2(1-\dbbb_i\bbb_i)\bbb_i\,,\\
    \sigmahat_i^{\widetilde{-}}&= 2\dbbb_i(1-\dbbb_i\bbb_i)\,,\\
    \sigmahat_i^\ztil &= 1-2\dbbb_i\bbb_i\,.
\eea

The bosonic modes are then approximated with a Gaussian ansatz, whose validity requires our above construction of the comoving frame: despite the exact spin-to-boson mapping~\eqref{eq:bosonize-sigma}, the resulting bosonic states are, in general, highly non-classical, due to the fact that the local spin basis states $\{\ket{\uparrow}_\ztil,\ket{\downarrow}_\ztil\}$ are mapped to Fock states $\{\ket{0},\ket{1}\}$. By aligning the $\ztil$ axis with the spin's orientation, we effectively redefine the vacuum of the bosonic mode, such that the remaining fluctuations can be approximated on the quadratic level with the Gaussian ansatz. A control parameter of the approximation can then be naturally defined as
\bea
    \epsilon \equiv \dfrac{1}{Ns}\sum_i\langle\dbbb_i\bbb_i\rangle=\dfrac{1}{N}\sum_i\left( 1 - \sqrt{\Sigma_{\tilde{\mu}}(\langle\shat_i^{\tilde{\mu}}\rangle/s)^2} \right)\,,
\eea
which measures the deviation from the bosonic vacuum as witnessed by the average number of excitations, i.e., the \textit{spin-wave density},
and can be regarded as the average deficit in the length of the normalized spin magnetization vector.

This Gaussian approximation allows us to uniquely specify the state of the entire system with the following $O(N^2)$ variational parameters:
    \begin{itemize}
        \item orientation of the comoving frames --- $\hat{Q}_i$\,,
        \item first moments --- $\beta_i\equiv\langle\bbb_i\rangle$\,,
        \item second moments --- $ u_{ij}\equiv \langle \delhat_i\delhat_j\rangle\,, v_{ij}\equiv \langle \ddelhat_i\delhat_j\rangle$\,,
    \end{itemize}
	where $\delhat_i\equiv\bbb_i-\beta_i$. Note that the construction of the comoving frame and the Gaussian approximation implies that $\beta_i=0$ for a \textit{static} state, yet this quantity plays a central role in determining the dynamics of the frame $\Qhat_i$, via the procedure we explain below.
    
     After initializing the variational parameters such that the frames $\Qhat_i$ are aligned with each spin (i.e., $\beta_i=0$), our algorithm can be conceptually understood as a stroboscopic procedure repeating the two following operations in every discretized time step $\delta t$, until the desired final time $t$ is reached:
\begin{enumerate}
		\item Calculate the infinitesimal increments for the Gaussian parameters $\delta\beta_i$, $\delta u_{ij}$, and $\delta v_{ij}$ using the stochastic master equation, and update these quantities with the increments. Note that the frame $\Qhat_i$ is kept constant within this step, which implies that the local $\ztil_i$ axes no longer align with the individual spins after the update as $\delta \beta_i\neq 0$ in general.
		
		\item Update local frames $\Qhat_i$ each along a \textit{geodesic} rotation [see Fig.~\ref{fig:spins} (c)] such that $\beta_{i}=0$ after the rotation. This can be achieved self-consistently by considering the evolution of the bosonic mode $\bbb_i$ generated by the (passive) rotation of the frame alone. Finally, increase the time by $\delta t$ and start a new iteration.
	\end{enumerate}

From the procedure described above, we note that the first moments $\beta_i$ are dummy variables since they are always zero at the end of each iteration, and are used only to compute the dynamics of the frame parameters $Q_i$. The first step is standard as in usual Gaussian approximation methods~\cite{verstraelen2018gaussian,liMonitoredLongrangeInteracting2025a},  where the expectation value of any higher moment is decomposed into first and second moments using the Wick theorem, allowing us to close the equations on the Gaussian level, and we derive their expressions in the Appendix~\ref{app:swqt-unravel-eom}. In what follows, we elaborate on the second step, which crucially depends on the parametrization of the frame $\Qhat_i$. We show that the quaternion formalism naturally provides a non-singular parametrization of the comoving frame, and the above procedure results in a stochastic differential equation for the variational parameters in the limit of $\delta t\to 0$.

\subsection{Non-singular parametrization of the comoving frame with quaternions}\label{sec:quat}

We now construct the comoving frame $O\xtil_i\ytil_i\ztil_i$. As illustrated in Fig.~\ref{fig:spins} (b), any local frame with $\ztil$ aligned with the spin vector is a valid choice, which leaves the orientation within the $O\xtil\ytil$ plane a gauge freedom with no physical consequence. Instead of fixing a gauge, which is equivalent to defining a unit vector field on $\mathbb{S}^2$ and will hence lead to discontinuity, we leave it as a dynamical degree of freedom that evolves naturally with the state. 

Let us consider the local frame of the $i-$th spin parametrized by the generic rotation
\bea
    \Qhat_i(\theta_i,\nvec_i)\equiv \rme^{-\rmi\theta_i\nvec_i\cdot\hat{\vec{s}}_i}\,,
\eea
and derive its dynamics that preserves the alignment with the local spin magnetization. In the rest of this derivation, we drop the site index $i$ for ease of notation, as the same procedure applies to each individual site $i$ independently. An efficient and non-singular way to encode this rotation is via the unit \textit{quaternion}
\bea
    Q\equiv\left(\cos\dfrac{\theta}{2},\nvec\sin\dfrac{\theta}{2}\right)\,,
\eea
which we also denote by $Q$ since they form a representation of the rotation group $SO(3)$ under quaternion multiplication. This is manifest from the representation of quaternions via Pauli matrices,
\bea
    q = (q_r,q_x,q_y,q_z) \mapsto q_r\idhat - \rmi \vec{q}\cdot\vec{\sigmahat}\,,
\eea
with the short-hand notation $\vec{q}\equiv(q_x,q_y,q_z)$. The unit quaternion $Q$ is then represented by
\bea
    Q \mapsto 
    \cos\dfrac{\theta}{2}\idhat - \rmi\sin\dfrac{\theta}{2}\nvec\cdot\vec{\sigmahat} = \rme^{-\rmi\frac{\theta}{2}\nvec\cdot\vec{\sigmahat}}\,.
\eea
In particular, the non-commutative quaternion product $Q_1Q_2$ corresponds to the rotation $\Qhat_1\Qhat_2$. 

To find the dynamics of $Q$, we proceed in the same way as in our previous work~\cite{liMonitoredLongrangeInteracting2025a}. After an infinitesimal time evolution of the state in the static frame $Q(t)$, we pause the physical dynamics to find the rotation $R(\d t): Q(t)\mapsto Q(t+\d t)$ that restores the alignment, as graphically represented in Fig.~\ref{fig:spins} (c). Suppose the bosonic mode's first moment $\langle\bbb\rangle$ evolved from 0 to $\beta$ and we parametrize the alignment rotation as $\Rhat(\phi,\mvec)=\rme^{-\rmi\phi\mvec\cdot\hat{\vec{s}}}$, for some angle $\phi$ and unit vector axis $\mvec$ to be determined. When the angle evolves with $\mvec$ fixed, the fictitious dynamics induced on the frame-dependent operator $\bbb(\phi)$ can be found by considering an inertial Hamiltonian $\Hhat_\rf$. Denoting $\Qhat(t+\d t)=\Rhat(\phi,\mvec)\Qhat(t)\equiv\Qhat(\phi)$, we have,
\bea
    \Hhat_\rf &= -\rmi\dfrac{\d\Qhat^\dagger(\phi)}{\d\phi}\Qhat(\phi) \\
    &= -\rmi\Qhat^\dagger(t)
    \dfrac{\d\Rhat^\dagger}{\d\phi}\Rhat\Qhat(t) \\&= \Qhat^\dagger(\phi)\left(\mvec\cdot\hat{\vec{s}}\right)\Qhat(\phi) \\&= \mvec\cdot\Tilde{\vec{s}}(\phi)\,,
\eea
where $\Tilde{\vec{s}}(\phi)$ is the vector of \textit{comoving} spin operators, which we now rewrite in terms of the bosonic operator $\bbb$ using the Holstein-Primakoff mapping~\eqref{eq:bosonization-n}. To obtain an integrable inertial Hamiltonian, we truncate the series at the \textit{lowest} order [resulting in $f(\nhat)\simeq 1$ in terms of Eq.~\eqref{eq:fn}] in this step, which is justified at low spin-wave densities.
This results in
\bea
    \Hhat_\rf &= \sqrt{\dfrac{s}{2}}(m_x+\rmi m_y)\dbbb + \sqrt{\dfrac{s}{2}}(m_x-\rmi m_y)\bbb \\&+ m_z(s-\dbbb\bbb)\,,
\eea
which is quadratic in $\bbb$.
It is then straightforward to solve for the dynamics of the comoving operator $\bbb$,
\bea
    \dfrac{\d\bbb}{\d\phi} = \rmi \left[ \Hhat_\rf, \bbb \right] = -\rmi\sqrt{\dfrac{s}{2}}(m_x+\rmi m_y) + \rmi m_z\bbb\,.
\eea
Our goal is to find $\mvec$ and $\phi$ as a function of $\beta$ such that $\langle\bbb(\phi)\rangle=0$. The simplest solution can be obtained by setting $m_z=0$, implying that the rotation axis is perpendicular to $\ztil$ axis, which corresponds to a geodesic rotation path passing through the north pole of the $Q(t)$ frame. It then follows that 
\bea
    \bbb(\phi) - \bbb(0) = -\rmi\sqrt{\dfrac{s}{2}}(m_x+\rmi m_y)\phi\,.
\eea
This shows that the rotating-frame evolution gives a displacement to the bosonic operator $\bbb$ (to the leading order of the Holstein-Primakoff transformation), and therefore the covariances are not affected by the rotation.
Our solution for the angle and the axis is therefore
\bea
    \phi = |\betatil|\,,\quad \mvec=\dfrac{1}{|\betatil|}({\Im\betatil}, -\Re\betatil, 0)\,,
\eea
with $\betatil\equiv\beta\sqrt{2/s}$. We have thus obtained the rotation $\Rhat$ as a function of $\beta$, which is, in quaternion form,
\bea\label{eq:Rbeta}
    R(\beta) &= \left(\cos\dfrac{\phi}{2},\mvec\sin\dfrac{\phi}{2}\right)\\&= \left( \cos\dfrac{|\betatil|}{2},\dfrac{1}{2}\Im\betatil\,\sinc\dfrac{|\betatil|}{2},-\dfrac{1}{2}\Re\betatil\,\sinc\dfrac{|\betatil|}{2},0 \right)\,,
\eea
where $\sinc(x)\equiv\sin (x)/x$ is the unnormalized sinc function. This is the rotation needed if $\betatil$ is a finite quantity, i.e.,
\bea\label{eq:dQ-finite}
    \d Q = (R-1)Q\,,
\eea
such as when a quantum jump occurs~\footnote{See Appendix~\ref{app:local-qj} for technical details in implementing local quantum jumps, where this result requires further correction.}. Otherwise, $\betatil$ is simply the infinitesimal $\d\betatil$,
and we can exploit the expansions
\bea
    \cos\dfrac{|\d\betatil|}{2} &= 1 - \dfrac{|\d\betatil|^2}{8} + O(\d t^2)\,,\\
    \sinc\dfrac{|\d\betatil|}{2} &= 1 +O(\d t)\,.
\eea
This gives us the form $R=1+\d A$ with
\bea
    \d A = \left(-\dfrac{|\d\betatil|^2}{8},\dfrac{1}{2}\Im\d\betatil,-\dfrac{1}{2}\Re\d\betatil,0\right)\,,
\eea
and the dynamical equation for the quaternion $Q(t)$ is simply,
\bea\label{eq:dQ}
    \d Q(t) = \d A(t) Q(t)\,.
\eea
Note that this equation is non-singular and can be readily cast into a stochastic differential equation. In particular, if the system dynamics is deterministic [such as in the mean-field limit~\cite{liEmergentDeterministicEntanglement2025} or during the non-Hermitian evolution of a quantum-jump unraveling (see Appendix~\ref{app:qj-eom})], we have $|\d\beta|^2=0$ and $\d Q(t)\propto \d t$, resulting in an ordinary differential equation.

\subsection{Summary of the method}

We now briefly recap the key operations performed in our algorithm. We first obtain the expressions of $\d\beta_i$, $\d u_{ij}$ and $\d v_{ij}$ using the stochastic master equation (see Appendix~\ref{app:swqt-unravel-eom} where we derive them for both the heterodyne and the quantum-jump unravelings), and then derive $\d Q_i$ from $\d \beta_i$ using Eq.~\eqref{eq:dQ} (for the continuous evolution) or Eq.~\eqref{eq:dQ-finite} (for the discontinuous jump), thus eliminating the dummy variables $\d\beta_i$ and arriving at a closed set of stochastic differential equations for $Q_i(t)$, $u_{ij}(t)$ and $v_{ij}(t)$, which fully determines the dynamics of the state within our semiclassical approximations.

Finally, let us remark that our construction of the \textit{local} spin-wave approximations and the inclusion of higher-order corrections in the bosonization significantly enlarges the domain of applicability of spin-wave quantum trajectories, beyond the conventional regime of large-$s$ or long-range interacting systems: in fact, it is exact in representing quantum trajectories of a single spin-$1/2$ (where the standard spin-wave theory completely breaks down), and provides a reasonably good approximation even at weakly coupled few-body systems with nearest-neighbor interactions [while the conventional spin-wave theory typically relies on the long-range (or high-dimensional) many-body limit], as we will benchmark in the next section.  Strengthened by the singularity-free parametrization via the quaternion formalism, our framework enables the efficient simulation of large-scale spin systems, which we now explore below.

\section{Results}\label{sec:results}

We now apply our generalized formalism of spin-wave quantum trajectories to the power-law spin models defined in Sec.~\ref{sec:long-range-local-diss-model}. We first present the results on Model I with the $\Hhat^{z-xx}$ Hamiltonian in Sec.~\ref{sec:result-zxx} below, where the heterodyne unraveling is considered. The results on Model II with $\Hhat^{x-zz}$ will be presented in Sec.~\ref{sec:qj-swqt-xzz}, where we employ the quantum-jump unraveling.

\subsection{Results on Model I with $\Hhat^{z-xx}$}\label{sec:result-zxx}

\begin{figure}[t]
    \centering
    \includegraphics[width=\linewidth]{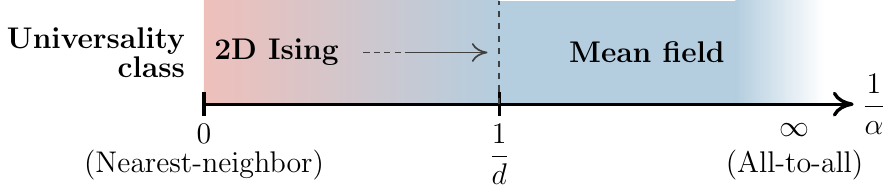}
    \caption{Pictorial summary of our main results on Model I (with the $\Hhat^{z-xx}$ Hamiltonian). The steady state displays a continuous phase transition associated with the spontaneous breaking of the $\mathbb{Z}_2$ symmetry, and the universality class exhibits a crossover from 2D Ising to mean field when the interaction range is increased from nearest-neighbor to long range.}
    \label{fig:result-sketch}
\end{figure}

Let us now consider Model I, where the Hamiltonian $\Hhat^{z-xx}$ as defined in Eq.~\eqref{eq:ham-zxx} has the drive along $z$ and two-body interaction along $x$, together with the local dissipation $\Lhat_i=\sqrt{\gamma}\sigmahat^-_i$ along the drive axis. The model hosts a continuous phase transition breaking the $\mathbb{Z}_2$ symmetry of the Liouvillian, and we study its critical behavior along the heterodyne trajectories. A sketch of our main findings is shown in Fig.~\ref{fig:result-sketch}, where we show that when the interaction range is increased from nearest-neighbor ($\alpha=\infty$) to long-range ($\alpha\leq d=2$), the universality class of the dissipative criticality displays a crossover from 2D Ising to mean field, witnessed by the critical exponents captured within our framework.

In the following, we first perform a simple benchmark (Sec.~\ref{sec:sw-het-bench}) showing that our construction of the generalized spin-wave approximations allows us to perfectly capture the trajectory dynamics in the weak-coupling regime. In Sec.~\ref{sec:ssb}, we discuss the single-trajectory behavior along the unraveled dynamics and explain the choice of the heterodyne unraveling for this model. Sec.~\ref{sec:ss-alpha1} presents a case study of the long-range interacting regime with $\alpha=1$, followed by the universality class crossover presented in Sec.~\ref{sec:universality}.

\subsubsection{Benchmark}\label{sec:sw-het-bench}

\begin{figure*}[t]
    \centering
    \includegraphics[width=0.95\linewidth]{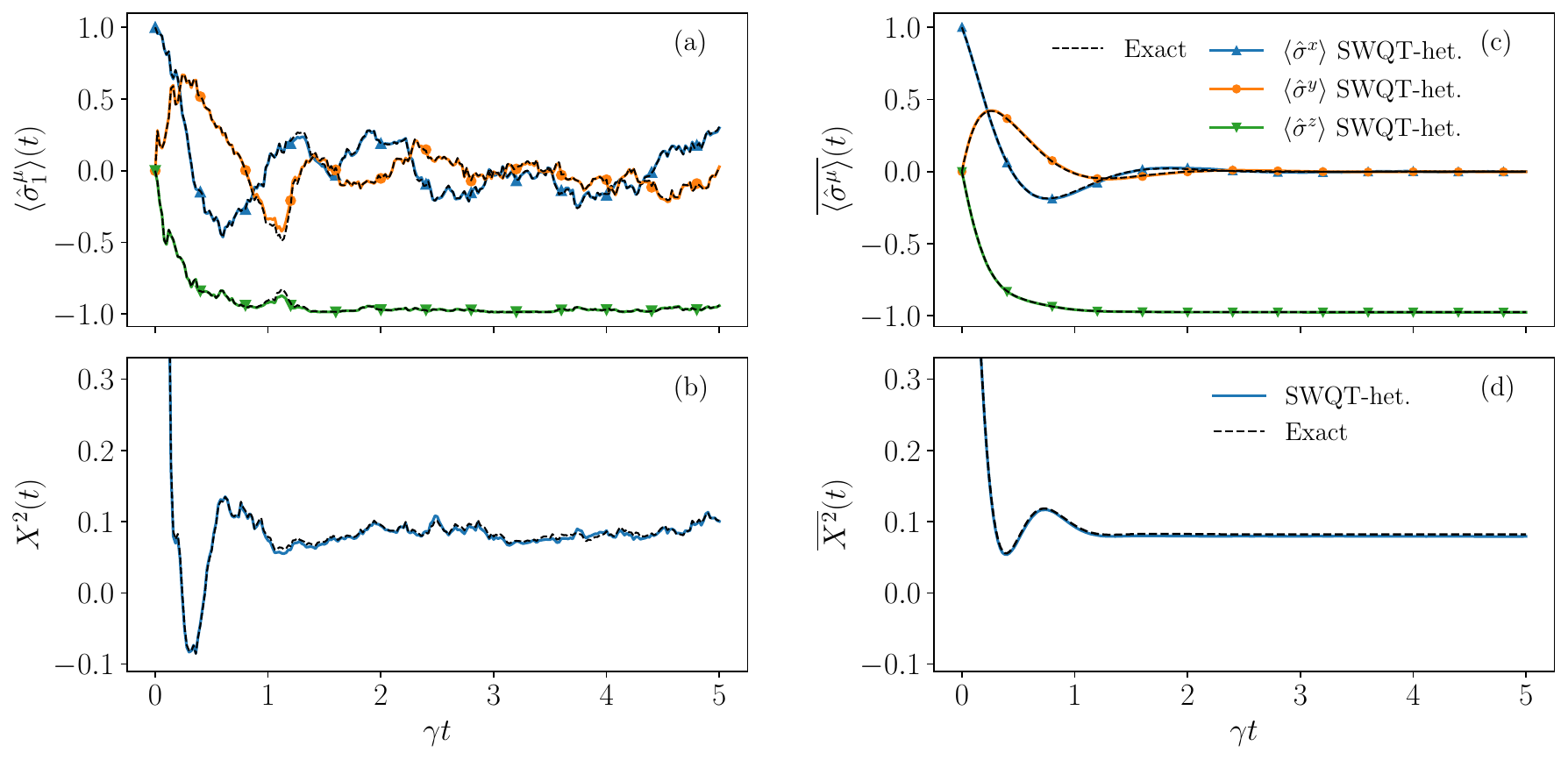}
    \caption{Benchmark of the spin-wave quantum trajectories (SWQT) with heterodyne (het.) unraveling for a $2\times 2$ spin system with $h=2\gamma$, $J=0.5\gamma$, and nearest-neighbor interactions ($\alpha=\infty$) for the following physical quantities: (a) single-trajectory expectation values on the first site $\langle\sigmahat^{x,y,z}_1\rangle$; (b) single-trajectory two-point correlation function $X^2$ [cf. Eq.~\eqref{eq:X2}]; (c) trajectory-averaged and site-averaged spin observables $\overline{\langle\sigmahat^{x,y,z}\rangle}$; (d) trajectory-averaged two-point correlation function $\overline{X^2}$. They are plotted as a function of time and shown together with the numerically exact solutions (see legend). The simulations are performed with a fixed time step $\gamma\delta t=10^{-4}$ and trajectory averages are performed over 1600 realizations.}
    \label{fig:swqt-het-bench}
\end{figure*}

To demonstrate the validity of the (unconventional and possibly counterintuitive) local spin-wave approximations, we first benchmark this method against exact solutions in a small system with $N=2\times 2$ spins at $h=2\gamma$, $J=0.5\gamma$ and nearest-neighbor interactions ($\alpha=\infty$), as shown in Fig.~\ref{fig:swqt-het-bench}, where both the spin-wave quantum trajectories (SWQT) and the exact ones are solved with the same fixed time step $\d t$ and the noise realization in the heterodyne dynamics. Fig.~\ref{fig:swqt-het-bench} (a) shows the fine agreement between the spin-wave results and the exact ones on the local spin observables $\langle\sigmahat^\mu_i\rangle$ of the site $i=1$ along a \textit{single trajectory}. We report that each site adopts a different magnetization direction along the trajectory evolution, due to the mutual independence of the heterodyne noise~\eqref{eq:dZ-het} in each local damping channel. This result signals, therefore, the success of the \textit{local} spin-wave approximations. Fig.~\ref{fig:swqt-het-bench} (b) shows the benchmark of the (site-averaged) two-point correlation function, defined as
\bea\label{eq:X2}
    X^2\equiv\sum_{i\neq j}\langle\sigmahat^x_i\sigmahat_j^x\rangle/N(N-1)\,,
\eea
evaluated along the same trajectory. Remarkably, the spin-wave quantum trajectory accurately captures the correlation dynamics in the presence of a weak coupling $J$. Fig.~\ref{fig:swqt-het-bench} (c) and (d) show, respectively, the trajectory-averaged magnetization $\overline{\langle\sigmahat^\mu\rangle}$ (averaged over all sites) and correlation function $\overline{X^2}$, showing again perfect agreement with the exact results.
We report that the average spin-wave density is $\overline{\epsilon}\lesssim0.02$, thereby validating the results.
Note that such exactitude should not be expected in general for an approximation method of semiclassical nature, and that it will eventually break down at higher couplings when the system inevitably enters the deep quantum regime. Nevertheless, we expect the generalized spin-wave approach to provide a plausible approximation capturing relevant physics well beyond the mean-field level, as suggested by this benchmark, while maintaining scalability to larger systems beyond the reach of exact methods: indeed, the Gaussian ansatz requires only $O(N^2)$ parameters to store the quaternions $Q_i$ and the covariances $u_{ij}$ and $v_{ij}$, and therefore, is extremely efficiently at capturing semiclassical dynamics.

\begin{figure}[h]
    \centering
    \includegraphics[width=\linewidth]{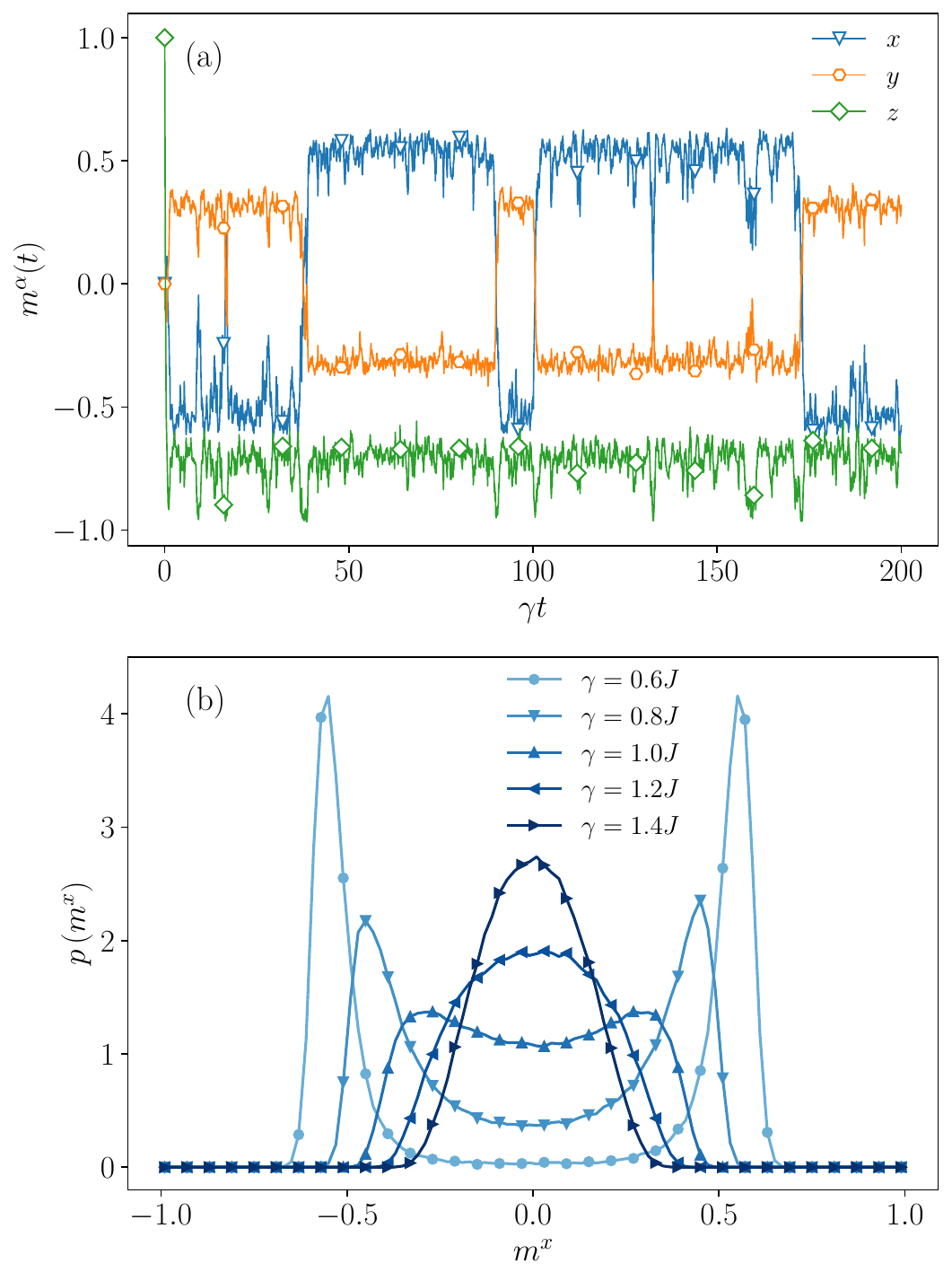}
    \caption{(a) SWQT heterodyne dynamics for a single trajectory with $N=6\times 6$, $h=J$, $\gamma=0.6J$ and $\alpha=1$ showing the symmetry-breaking behavior. (b) Distribution of $m^x$ along trajectories for different values of $\gamma$ (see legend) showing the bimodality of the trajectory ensemble.}
    \label{fig:bimod}
\end{figure}

\subsubsection{Symmetry breaking along trajectories}\label{sec:ssb}

Let us now apply the generalized SWQT framework to larger system sizes and study the temporal trajectory dynamics in the power-law spin system. As described in Sec.~\ref{sec:long-range-local-diss-model}, the Liouvillian of the model I admits a $\mathbb{Z}_2$ symmetry due to its invariance under the transformation $\shat^{x,y}_i \mapsto -\shat^{x,y}$. This implies that at any finite sizes, the (unique) steady state exhibits the same symmetry and therefore always has $m^{x,y}=0$, where we define the notation for the average magnetization $m^\mu\equiv\sum_i\langle\sigmahat^\mu_i\rangle/N=0$. In contrast, the stochastic master equation for the heterodyne unraveling~\eqref{eq:het-drho} does not share this symmetry due to the presence of terms of the form $\Lhat_i\rhohat$ and $\rhohat\dLhat_i$, thus offering a perfect playground for observing the symmetry-breaking dynamics in real time. Fig.~\ref{fig:bimod} (a) shows a single SWQT heterodyne trajectory for $N=6\times 6$ with $h=J$, $\gamma=0.6J$ and $\alpha=1$ corresponding to the mean-field symmetry-broken regime, where the $x-$ and $y-$ site-averaged spin components switch between two metastable branches, with a switching time on the order of the inverse Liouvillian gap~\cite{mingantiOutofEquilibriumPhaseTransitions2018}. 
Fig.~\ref{fig:bimod} (b) shows the distribution of $m^x$ at long times along trajectories of different values of $\gamma$, revealing the bimodality of the trajectory ensemble: the average state can be viewed as a symmetric mixture of states with opposite $x-$ (and $y-$) magnetization in the ferromagnetic phase. This bimodality disappears when the system enters the disordered phase with larger values of $\gamma$. 

A few remarks are in order. The above bimodality implies that the average mixed state is highly non-Gaussian and therefore is very difficult to capture with semiclassical approximations operating on the level of the mixed density matrix. The SWQT framework successfully reproduces this essential feature in symmetry-breaking phase transitions, thanks to two crucial facts, that 1) the stochastic sampling of spin-wave trajectories effectively constructs a mixture of Gaussian states, resulting in a larger family of states that can be reached on the level of the average density matrix, and that 2) the choice of the heterodyne unraveling explicitly breaks the symmetry on the level of the quantum trajectory, such that our semiclassical state representation does not need to struggle in trying to approximate a symmetric bimodal state. This will no longer be the case if one chooses a symmetry-preserving trajectory dynamics, such as the quantum-jump unraveling (since $\Lhat_i$ and $\dLhat_i$ always appear together, and that symmetry transformation simply flips the sign of $\Lhat_i=\sqrt{\gamma}\sigmahat^-_i$, leaving the stochastic equation invariant). The true state along the quantum-jump dynamics therefore remains (or quickly relaxes to be) already bimodal and resembles a \schr{} cat state~\cite{mingantiOutofEquilibriumPhaseTransitions2018} (as it is also a pure state, i.e., the bimodality must result from a coherent superposition), which is highly quantum and falls beyond the reach of semiclassical approximations. Nevertheless, our spin-wave framework can be applied to the quantum-jump unraveling in the absence of this symmetry, which we demonstrate on the $\Hhat^{x-zz}$ model in Sec.~\ref{sec:qj-swqt-xzz}.

\begin{figure}[t]
    \centering
    \includegraphics[width=0.86\linewidth]{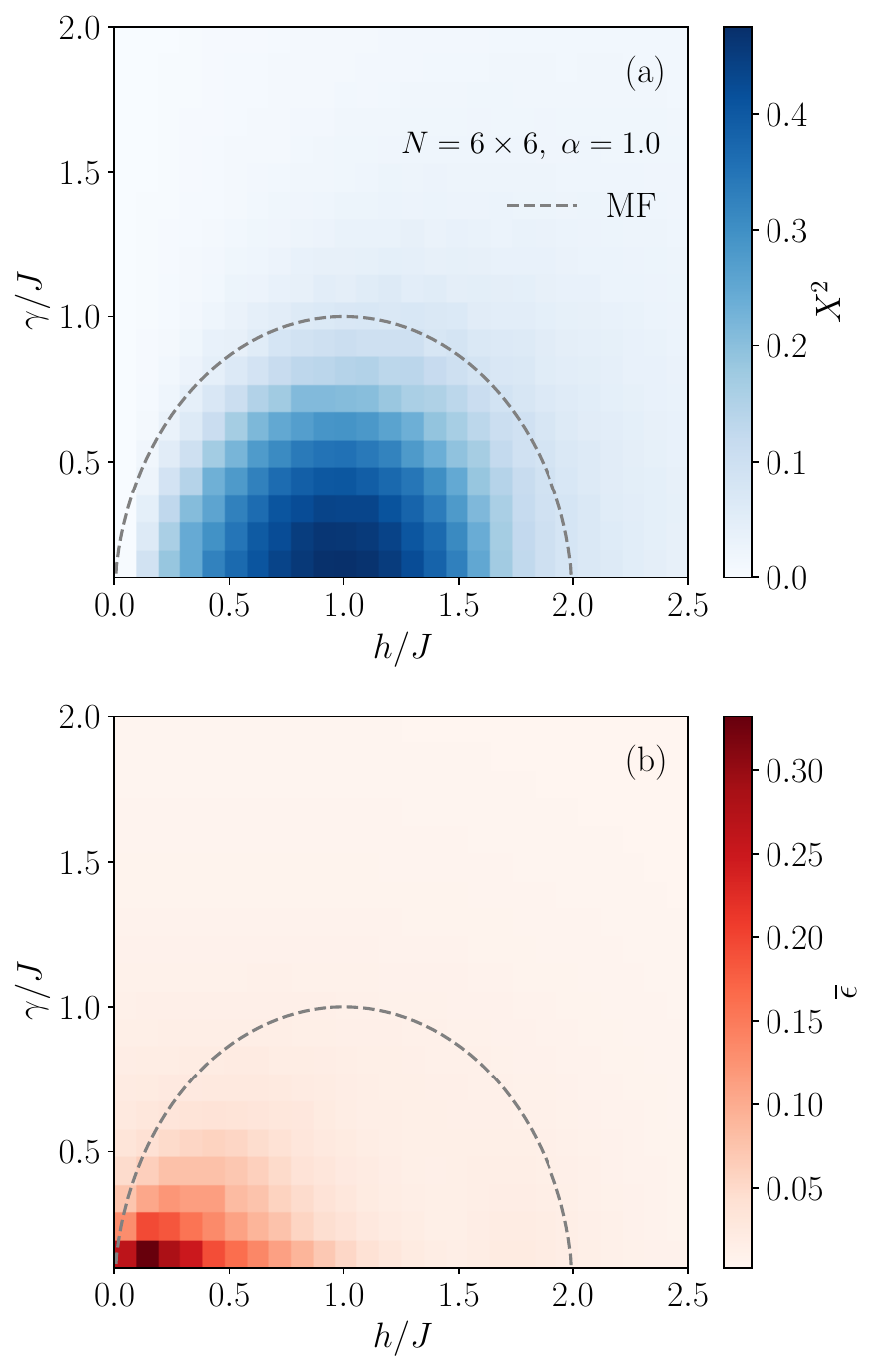}
    \caption{Steady-state results for a system with $N=6\times 6$, $\alpha=1$ in the steady state, obtained with spin-wave quantum trajectories. (a) Magnetization order parameter $X^2$ as a function of the external drive $h$ and the dissipation $\gamma$. (b) Trajectory-averaged spin-wave density $\overline{\epsilon}$. In both panels, the mean-field (MF) phase boundary is marked with a dashed line (see legend). }
    \label{fig:x2_ent_heatmap_6}
\end{figure}

\subsubsection{Steady-state phases with long-range interactions $\alpha=1$}\label{sec:ss-alpha1}
The dynamics presented in the previous section suggests the onset of a symmetry-breaking phase transition in the steady-state of the dynamics. To probe the order-disorder crossover at finite $N$, we adopt the two-point correlation function $X^2$ defined in Eq.~\eqref{eq:X2} as the order parameter, which effectively measures the second moment of the distribution of $m^x$. In what follows, we will refer to $X^2$ as the ``magnetization'' when there is no ambiguity (since in the mean-field limit it corresponds to the squared $x-$ magnetization). Fig.~\ref{fig:x2_ent_heatmap_6} shows the steady-state results for a $N=6\times 6$ lattice with interaction strength $\alpha=1$. Panel (a) shows the magnetization $X^2$, which presents a region with a nonzero value, aligning well with the mean-field prediction. This is expected as the interaction range is long enough ($\alpha<d$ with $d=2$ the dimensionality), such that the mean-field theory is exact in the thermodynamic limit~\cite{paz2021}. Fig.~\ref{fig:x2_ent_heatmap_6} (b) shows the trajectory-averaged spin-wave density $\overline{\epsilon}$, which displays a maximum at low drive $h$ and interaction $J$, within the mean-field ordered phase. This suggests the proliferation of quantum fluctuations, where the results should be regarded as qualitative, as the elevated spin-wave density signals less accurate approximations.

\begin{figure}[t]
    \centering
    \includegraphics[width=\linewidth]{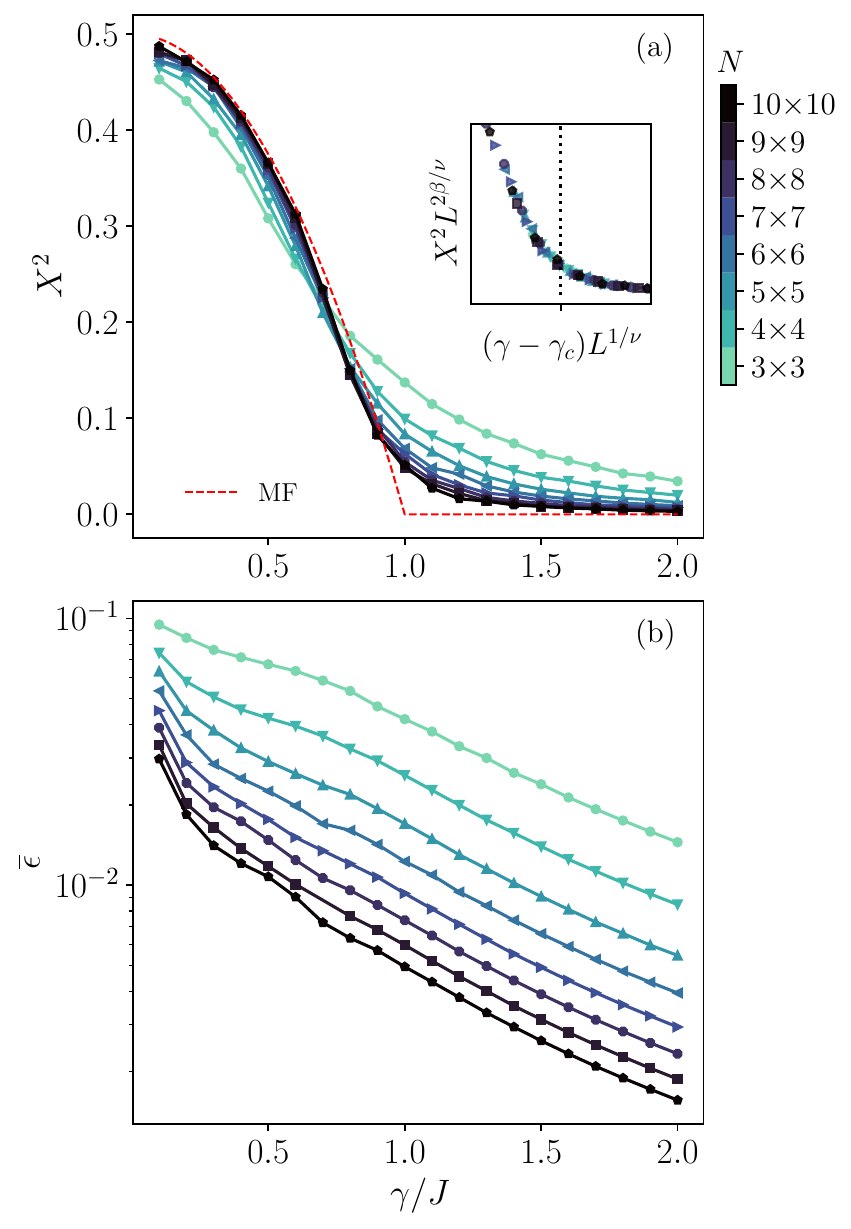}
    \caption{Steady-state results for the same quantities as in Fig.~\ref{fig:x2_ent_heatmap_6} with $\alpha=1$ and $h=J$, as a function of the dissipation $\gamma$ and system size $N=L\times L$ (legend shared across both panels). (a) Magnetization order parameter $X^2$ shown together with the mean-field (MF) result. Inset: finite-size scaling of the results with $\gamma_c/J = 1.02 \pm 0.07$, $\beta = 0.46 \pm 0.10$, $\nu = 1.00 \pm 0.05$ extracted from a collapsing analysis, with the vertical dotted line marking the extracted critical point.  (b) Trajectory-averaged spin-wave density $\overline{\epsilon}$.}
    \label{fig:ss-alpha1-x2-ent-sw}
\end{figure}

The same physical quantities are shown in Fig.~\ref{fig:ss-alpha1-x2-ent-sw} for fixed $h=J$, as a function of the dissipation $\gamma$ for different system sizes up to $N=10\times 10$. The magnetization order parameter $X^2$ asymptotically approaches the mean-field result, where the latter predicts a critical point at $\gamma=J$. This is confirmed with a finite-size scaling analysis, where the data for different sizes are fitted with the scaling ansatz
\bea\label{eq:scaling-ansatz}
    X^2(\gamma, L) = L^{-2\beta/\nu} \tilde{f}\left( (\gamma - \gamma_c) L^{1/\nu} /J\right)\,,
\eea
resulting in the curves collapsing when plotting $X^2 L^{2\beta/\nu}$ vs. $(\gamma-\gamma_c)L^{1/\nu}/J$, as shown in the inset of Fig.~\ref{fig:ss-alpha1-x2-ent-sw} (a). The critical point extracted from the fit is $\gamma_c = 1.02 \pm 0.07$, together with the magnetization critical exponent $\beta = 0.46 \pm 0.10$ and the correlation exponent $\nu = 1.00 \pm 0.05$, which matches well the mean-field prediction as well as the mean-field Ising universality class ($\beta=1/2$). Note that here we obtain $\nu\simeq 1$ instead of the ``standard'' value $1/2$ derived for an infinite system, which is a typical result in finite-size scaling for long-range interacting systems, where the hyperscaling relation breaks down, and one should expect instead $\nu=d/2$~\cite{botetLargesizeCriticalBehavior1983,binderFinitesizeTestsHyperscaling1985}. Finally, Fig.~\ref{fig:ss-alpha1-x2-ent-sw} (b) shows the average spin-wave density $\overline{\epsilon}$, which decays at higher dissipation $\gamma$ and with increasing system size. The latter is a feature of long-range interacting systems with $\alpha<d$, and this behavior witnesses the asymptotic exactness of the semiclassical approximation.

\begin{figure}
    \centering
    \includegraphics[width=0.8 \linewidth]{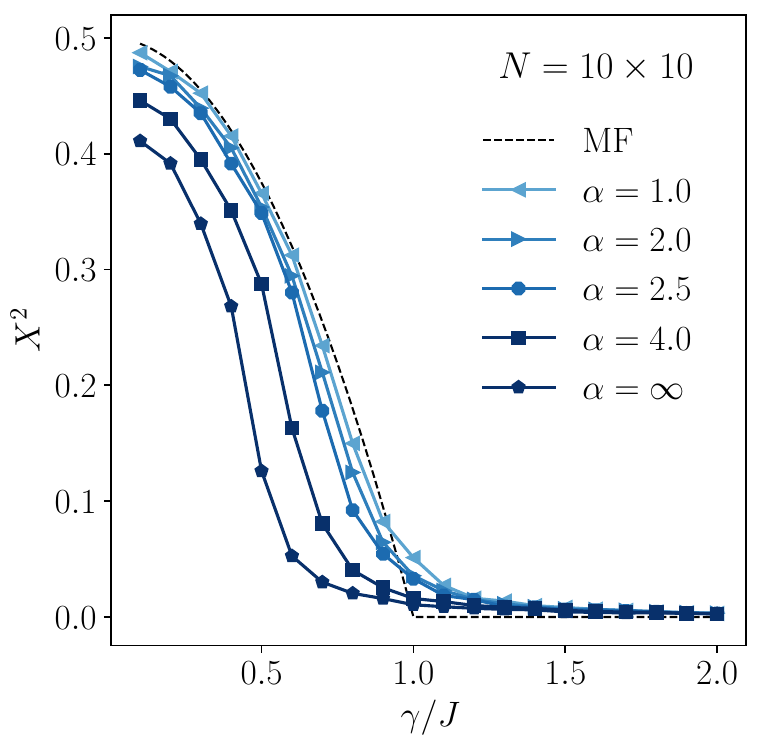}
    \caption{Steady-state magnetization $X^2$ as a function of $\gamma$ in a $10\times 10$ lattice, for different interaction ranges represented by $\alpha$, shown together with the mean-field result (see legend).}
    \label{fig:L10_x2_alp}
\end{figure}

\subsubsection{Universality class crossover}\label{sec:universality}

One of the advantages of the generalized SWQT framework is that its applicability reaches beyond the usual long-range ($\alpha<d$) regime typically required by the conventional spin-wave theory, and we now embark on the investigation of the effect of different interaction ranges on the symmetry-breaking phase transition.

\begin{figure}[t]
    \centering
    \includegraphics[width=\linewidth]{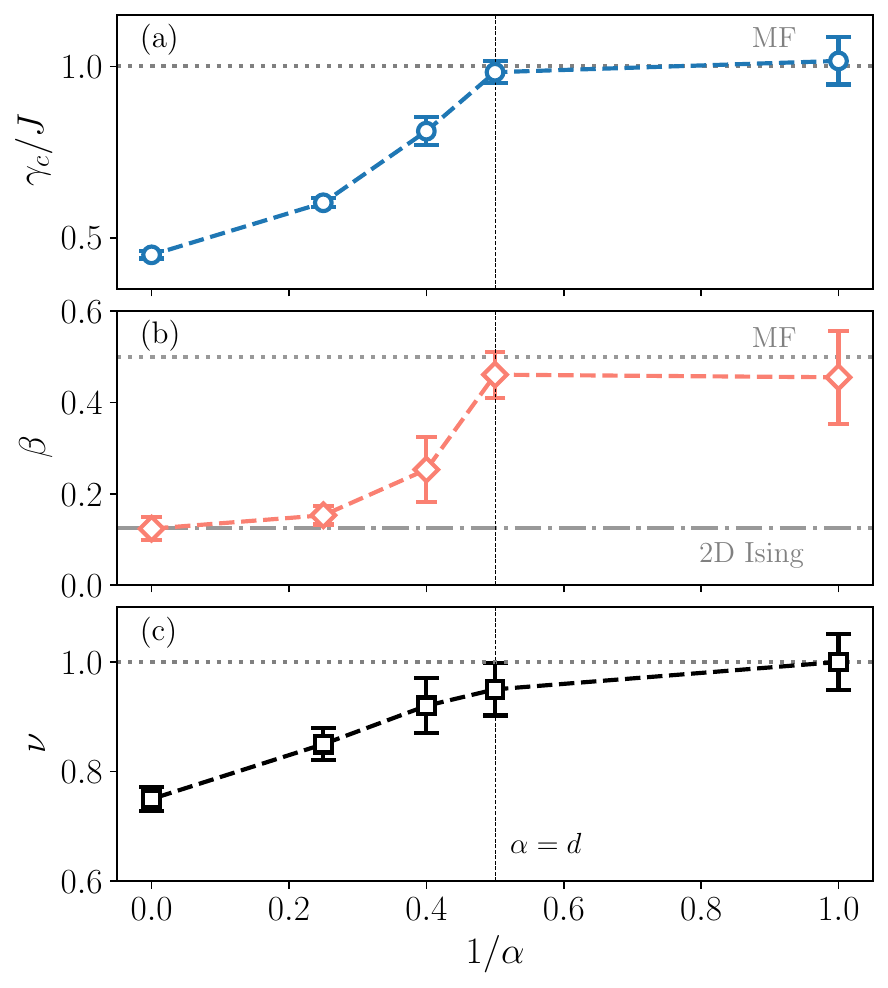}
    \caption{Finite-size scaling parameters as a function of interaction range represented by $1/\alpha$ showing the crossover of universality class, where the vertical dotted line marks $\alpha=d=2$. (a) Critical point $\gamma_c$, with the mean-field (MF) value $\gamma_c/J=1$ marked with a horizontal dotted line. (b) Magnetization critical exponent $\beta$, with the mean-field value $\beta=1/2$ marked with a horizontal dotted line and the value for 2D Ising universality class $\beta=1/8$ marked with a dash-dotted line. (c) Correlation length critical exponent $\nu$ with the horizontal dotted line marking the value $\nu=1$ for both the mean-field and the 2D Ising universality class. }
    \label{fig:FSS-vs-ak}
\end{figure}

At shorter interaction ranges ($\alpha>d$), one generally expects the system behavior to deviate from the mean-field universality class (when $d$ is below the upper critical dimension~\cite{landauStatisticalPhysics1980}) due to increased fluctuations. On the level of our semiclassical spin-wave theory, this is manifest via a non-vanishing spin-wave density providing a correction to the classical (zeroth-order) component of the theory. The most evident effect is a shifted critical point, which can already be observed at finite sizes. Fig.~\ref{fig:L10_x2_alp} shows the order parameter $X^2$ as a function of $\gamma$ in a $10\times 10$ lattice, with different interaction exponents $\alpha$ ranging from $\alpha=1<d$ (long-range) to $\alpha=\infty$ (nearest-neighbor interactions). As the interaction range decreases, the collective ordering becomes less robust, and we observe accordingly the curves deviating from the mean-field prediction, with the ordered region shrinking to smaller values of dissipation $\gamma$. This observation is further supported by finite-size scaling analyses [in the same way as demonstrated above in Fig.~\ref{fig:ss-alpha1-x2-ent-sw}], for all interaction ranges considered in Fig.~\ref{fig:L10_x2_alp} and for system sizes ranging from $N=3\times 3$ to $N=10\times 10$. The result is shown in Fig.~\ref{fig:FSS-vs-ak}. Panel (a) shows the critical point $\gamma_c$ as a function of the interaction range represented by $1/\alpha$. At long ranges ($\alpha\leq 2$), the critical point remains that of the mean-field value $\gamma_c=1$, while its value decreases with shorter ranges (smaller $1/\alpha$), yet saturates to a nonzero value of $\gamma_c/J=0.45\pm0.01$ at $\alpha=\infty$, i.e., in the presence of nearest-neighbor-only interactions. This suggests that the considered dimension $d=2$ is not below the lower critical dimension of the system, such that the ferromagnetic ordering survives the short-range fluctuations. Fig.~\ref{fig:FSS-vs-ak} (b) shows the magnetization critical exponent $\beta$ vs. the interaction range, which confirms our observations above. In particular, the extract value at $\alpha\leq 2$ matches that of the mean-field Ising universality class ($\beta=1/2$), and the result at $\alpha=\infty$ accurately coincides with $\beta=1/8$, which is the value for 2D Ising universality class, as expected from field-theoretic analysis~\cite{maghrebiNonequilibriumManybodySteady2016}, with intermediate values of $\beta$ interpolating between the two limits. This is a significant result highlighting the strength of the generalized SWQT framework, which predicts a crossover of the universality class tuned by the interaction range, and also captures the \textit{non-mean-field} critical exponent of the 2D Ising universality class in the short-range limit. Fig.~\ref{fig:FSS-vs-ak} (c)  displays the fitted correlation length exponent $\nu$ for different interaction ranges. As previously discussed in Sec.~\ref{sec:ss-alpha1}, the long-range regime of $\alpha<d$ implies the breakdown of hyperscaling relations leading to $\nu=d/2=1$, which is consistent with our finite-size scaling results. At short ranges, our extracted value gradually decreases, with its value at $\alpha=\infty$ being $\nu=0.75\pm 0.02$, deviating from that of the 2D Ising universality class ($\nu=1$). This underestimation indicates the limitations of the semiclassical approximations in fully capturing the diverging correlation length of the 2D criticality dominated by fluctuations.

\subsection{Results on Model II with $\Hhat^{x-zz}$}\label{sec:qj-swqt-xzz}

We discuss in this section the results on Model II, with the $\Hhat^{x-zz}$ Hamiltonian defined in Eq.~\eqref{eq:ham-xzz}, where the drive is along $x$, while the interaction is along $z$, the same axis along which the dissipation $\Lhat_i=\sqrt{\gamma}\sigmahat_i^-$ acts. Unlike the $\Hhat^{z-xx}$ model, the Liouvillian and its unraveled trajectory dynamics no longer have the $\mathbb{Z}_2$ symmetry.
For demonstration purposes, we show in this section the results obtained with the quantum-jump unraveling, and consider the 2D spin lattice with only nearest-neighbor interactions ($\alpha=\infty$), a regime combining ``worst-case scenarios'' for a standard spin-wave theory, that is also extensively adopted for benchmarking neural-network-based variational methods for spins~\cite{vicentiniVariationalNeuralNetworkAnsatz2019,Luo2022,Mellak2024}, where even state-of-the-art neural network state ans\"{a}tz struggle to capture its non-equilibrium steady-state.

\begin{figure}[h]
    \centering
    \includegraphics[width=0.9\linewidth]{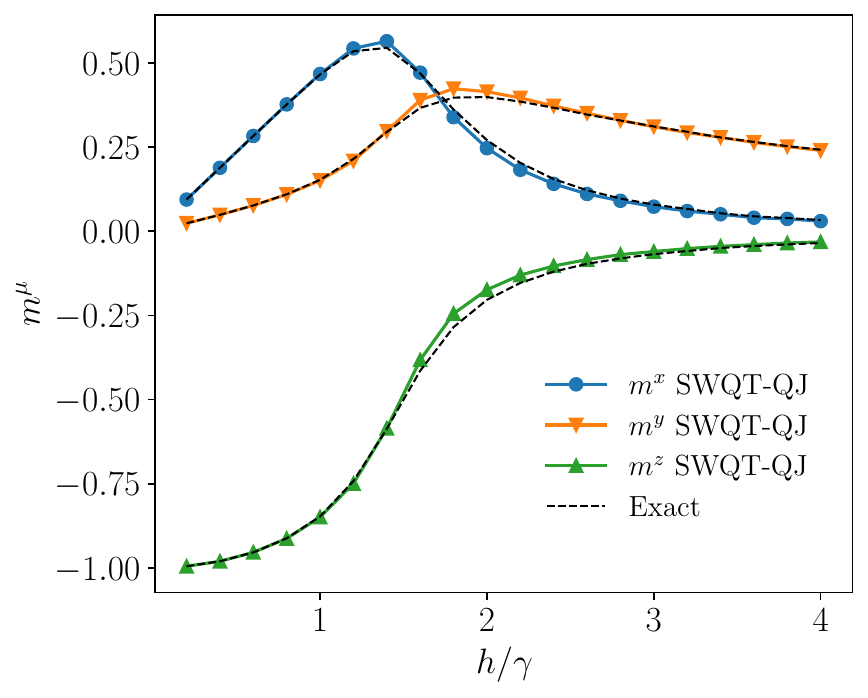}
    \caption{SWQT quantum jump (QJ) benchmark results for a $N=4\times 4$ spin lattice with $J=2\gamma$ and $\alpha=\infty$ : steady-state average magnetization $m^{x,y,z}$ obtained using SWQT with the QJ unraveling, together with the exact solution (see legend), as a function of the external field $h$.}
    \label{fig:swqj-bench}
\end{figure}

As a proof of principle, we perform a first demonstration of spin-wave theory applied to quantum-jump trajectories, and benchmark the results against the exact solution for a $N=4\times 4$ spin lattice with $J=2\gamma$, which is the same parameter regime considered in~\cite{vicentiniVariationalNeuralNetworkAnsatz2019,Luo2022,Mellak2024} for benchmarking neural-network quantum states. As shown in Fig.~\ref{fig:swqj-bench}, the SWQT approach captures surprisingly well the steady-state magnetization observables $m^{x,y,z}$, despite the simple (as compared to sophisticated neural-network constructions) Gaussian ansatz adopted. We report that the heterodyne SWQT produces almost identical results for this benchmark, yet the quantum-jump unraveling, when applicable, is much more numerically efficient as one can adopt standard ordinary differential equation solvers for the smooth dynamics between jumps (see Appendix~\ref{app:qj-eom}) instead of using a fixed time step.

\begin{figure}[h]
    \centering
    \hspace{0.17cm}\includegraphics[width=0.885\linewidth]{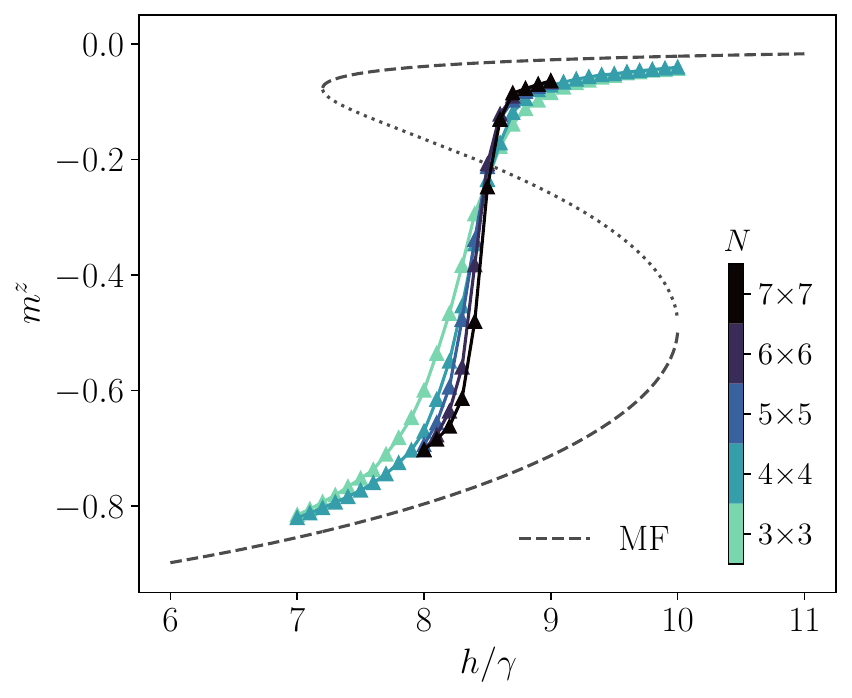}
    \caption{SWQT quantum jump (QJ) results for a 2D spin lattice with $J=7\gamma$ and $\alpha=\infty$ : steady-state average magnetization $m^{z}$ as a function of the external field $h$ for different system sizes, shown together with the mean-field solution (see legend).}
    \label{fig:xzz-qj-v3}
\end{figure}

Finally, Fig.~\ref{fig:xzz-qj-v3} shows the results at a higher interaction strength, $J=7\gamma$. In this regime, the mean-field theory predicts a bistable solution for the steady state, as shown by the S-shaped curve, where for drive values in the range $7.2\lesssim h/\gamma\lesssim 10$, three solutions of the steady-state are present: the lower and the higher branches are the stable solutions while the middle branch is unstable, analogous to the phenomenology of optical bistability~\cite{Vogel1989}. Exploiting the quantum-jump spin-wave trajectories, we obtain the steady-state magnetization $m^z$ as a function of the drive $h/\gamma$ deep inside the mean-field bistability region for different system sizes. By virtue of the beyond-mean-field quantum correlations included in our framework, the spin-wave results are immune to the mean-field hysteresis artifact and instead always capture a unique steady state. As the system size increases, the crossover from the lower branch to the higher one steepens, and the curves for different sizes intersect at $h_c\simeq 8.5\gamma$, suggesting the emergence of a first-order (discontinuous) dissipative phase transition at the critical point $h_c$, which is consistent with the results reported in the previous study~\cite{jinPhaseDiagramDissipative2018} employing a cluster mean-field approach on the same model.

\section{Conclusion}\label{sec:conclusions}

	In this work, we have proposed a semiclassical approximation framework based on a generalized stochastic spin-wave theory along quantum trajectories of driven-dissipative spin systems. {Our results highlight several features of the methodology that can be summarized as follows:
		\begin{itemize}
			\item The method significantly widens the applicability of spin-wave approximations, via the construction of local comoving frames, the inclusion of higher-order corrections, and the quaternion parametrization of the frames, which also completely resolved the coordinate singularity issues commonly present in time-dependent spin-wave theories.			
			\item Applying this framework to the 2D variable-range driven-dissipative spin model with a spontaneous $\mathbb{Z}_2$ symmetry-breaking phase transition, we unveil a crossover of universality class from mean-field to 2D Ising when the interaction range shortens. Remarkably, the method accurately recovers the non-mean-field critical exponent $\beta=1/8$ in the nearest-neighbor interacting limit. 
            
			\item The spin-wave framework is further generalized to incorporate local quantum-jump dynamics, with careful benchmarks showing its accuracy, and is showcased by capturing a first-order transition in a 2D nearest-neighbor interacting spin lattice without $\mathbb{Z}_2$ symmetry. 
		\end{itemize}
	}
	
	Beyond these results demonstrating the broad applicability of our method, several exciting directions are already manifest for future explorations. For example, the framework can be readily extended to include bosonic modes interacting with the spins, where the same Gaussian approximation is applied to the bosons, as demonstrated in our previous work~\cite{liMonitoredLongrangeInteracting2025a}. This allows us to explore a large class of light-matter interaction models, such as the Tavis-Cummings model~\cite{breuer2002theory,delmonteMeasurementinducedPhaseTransitions2025}, highly relevant for current synthetic quantum platforms.
	Moreover, as our method construction is insensitive to the spatial geometry of the physical system, we expect it to work equally well in any dimension, allowing us to probe the interplay between the interaction range and the dimensionality in the critical dynamics, which is a fundamental problem in statistical physics.
    Furthermore, the framework can be readily combined with the recently developed variational phase-space approaches~\cite{toscaEfficientVariationalDynamics2025,toscaVariationalDynamicsOpen2026}, to provide a beyond-semiclassical description of monitored dynamics in the deep-quantum regime. Finally, given the generality of our formalism, the proposed SWQT framework sets a natural and scalable playground for investigating the rich physics in open many-body spin systems, paving the way to the exploration of novel collective phenomena in non-equilibrium quantum matter.

\section*{ Acknowledgments}
	We would like to thank Jacopo Tosca, Alessio Paviglianiti and Giulia Salatino for insightful discussions. This work was supported by PNRR MUR project PE0000023- NQSTI, by the European Union (ERC, RAVE, 101053159). AD acknowledges funding from the European Research Council (ERC) under the European Union’s Horizon 2020 research and innovation programme (Grant agreement No. 101002955 — CONQUER). Views and opinions expressed are, however, those of the author(s) only and do not necessarily reflect those of the European Union or the European Research Council. Neither the European  Union nor the granting authority can be held responsible for them.

\appendix

\section{Mean-field equations for the spin models}\label{app:mf}

We derive in this appendix the mean-field equations for the spin models considered in this work. Consider the following form of the Hamiltonian,
\bea
    \Hhat = h\sum_i\sigmahat^{\nvec_1}_i + \sum_{i\neq j}J_{ij}^{(\alpha)}\sigmahat_i^{\nvec_2}\sigmahat_j^{\nvec_2}\,,
\eea
with $\sigmahat^{\nvec}\equiv \nvec\cdot\vec\sigmahat$, and the Lindblad jump operator $\Lhat_i=\sqrt{\gamma}\sigmahat^-_i$, which covers both models defined in the main text. In the mean-field approximation, the system state is described by a factorized ansatz,
\bea
    \rhohat^{\mathrm{MF}} = \bigotimes_i\rhohat_i\,,
\eea
and we further impose $\rhohat_i=\rhohat$, $\forall i$, due to the translational symmetry of the Liouvillian. The dynamics can then be solved in terms of the single-body state $\rhohat$,
\bea
    \dfrac{\d\rhohat}{\d t} &= -\rmi[\Hhat^{\mathrm{MF}}(\rhohat),\rhohat] 
    + \gamma\left(\sigmahat^+\rhohat\sigmahat^- - \dfrac{1}{2}\left\{ \sigmahat^+\sigmahat^-,\rhohat \right\}  \right)\\
    &=\lcal^{\mathrm{MF}}\rhohat\,,
\eea
where the mean-field Hamiltonian is
\bea
    \Hhat^{\mathrm{MF}}(\rhohat) = h\sigmahat^{\nvec_1} + 2J\mathrm{Tr}[\rhohat\sigmahat^{\nvec_2}]\sigmahat^{\nvec_2}\,,
\eea
and we recall that $\sum_{j\neq i_0} J_{i_0,j}^{(\alpha)}=J$, $\forall i_0$, due to the Kac normalization. 
In particular, the mean-field steady state can be obtained by solving $\lcal^{\mathrm{MF}}\rhohat=0$. The above derivation shows that the mean-field approximation ignores any quantum correlation between different sites, and that it is blind to the interaction range $\alpha$.

\section{Unravelling-specific spin-wave equations}\label{app:swqt-unravel-eom}

In this appendix, we derive the spin-wave equations for the two unravelings considered in the main text.

\subsection{Heterodyne dynamics}\label{app:het-eom}

Within the Gaussian approximation, the heterodyne equation~\eqref{eq:het-drho} translates into the stochastic dynamics of the variational parameters, which are conveniently described by the adjoint stochastic master equation. 
    For a time-independent operator $\Ohat$, its expectation value along the heterodyne dynamics evolves as follows,
\bea\label{eq:do-adjoint}
	\d\langle\Ohat\rangle = \d t \langle\Lcal^\ddagger\Ohat\rangle+\langle\d\hcal^\ddagger\Ohat\rangle\,,
\eea
where the adjoint superoperators are defined as follows,
\bea\label{eq:LO-HO-adjoint}
\lcal^\ddagger\Ohat\equiv&~\rmi[\Hhat,\Ohat] + \sum_i\left( \Lhat^\dagger_i\Ohat\Lhat_i-\dfrac{1}{2}\left\{ \Lhat^\dagger_i\Lhat_i,\Ohat \right\} \right)\,,\\
\d\hcal^\ddagger\Ohat\equiv&~ \sum_i\left[\d Z_i^*\Ohat\left(\Lhat_i-\langle\Lhat_i\rangle\right) +\right.\\  &\phantom{====}\left.\d  Z_i\left(\Lhat_i^\dagger-\langle\Lhat_i^\dagger\rangle\right)\Ohat\right]\,.
\eea

The first moment $\beta_i=\langle\bbb_i\rangle$ is the expectation value of a time-independent operator (since the frame remains static in this step), hence one can use directly the adjoint stochastic master equation~\eqref{eq:do-adjoint} to obtain
\bea
    \d\beta_i = \d t\langle \lcal^\ddagger\bbb_i\rangle + \langle\d\hcal^\ddagger\bbb_i\rangle\,.
\eea

The second moments are expectations of time-dependent operators by definition, of the form $\delhat_A\delhat_B=\Ahat\Bhat-\langle\Ahat\rangle\langle\Bhat\rangle$ for some time-independent operators $\Ahat$ and $\Bhat$, and their increments should be evaluated using stochastic differentiation rules. Let us first note that the adjoint superoperator $\lcal^\ddagger$ is linear, i.e.
\bea
    \lcal^\ddagger(\Ahat+\lambda\Bhat) = \lcal^\ddagger\Ahat+\lambda\lcal^\ddagger\Bhat\,,
\eea
and that adding multiples of identity does not contribute to the expectation value, i.e., $\langle\lcal^\ddagger(c\idhat)\rangle=0$, $\forall \lambda, c\in\mathbb{C}$, and the exact same properties are satisfied by the superoperator $\d\hcal$. This allows us to derive,
\bea
    \d\langle\delhat_A\delhat_B\rangle &= \d\langle\Ahat\Bhat\rangle-\d(\langle\Ahat\rangle\langle\Bhat\rangle)\\
    &= \d\langle\Ahat\Bhat\rangle - (\d\langle\Ahat\rangle)\langle\Bhat\rangle - \langle\Ahat\rangle\d\langle\Bhat\rangle - \d\langle\Ahat\rangle\d\langle\Bhat\rangle\\
    &= \langle(\lcal^\ddagger+\d\hcal^\ddagger)(\Ahat\Bhat)\rangle - \langle(\lcal^\ddagger+\d\hcal^\ddagger)\Ahat\rangle\langle\Bhat\rangle\\
    &- \langle\Ahat\rangle\langle(\lcal^\ddagger+\d\hcal^\ddagger)\Bhat\rangle - \underbrace{\langle(\lcal^\ddagger+\d\hcal^\ddagger)(\langle\Ahat\rangle\langle\Bhat\rangle)\idhat\rangle}_{=0}\\
    &- \langle\d\hcal^\ddagger\Ahat\rangle\langle\d\hcal^\ddagger\Bhat\rangle\\
    &= \langle(\lcal^\ddagger+\d\hcal^\ddagger)(\delhat_A\delhat_B)) - \langle\d\hcal^\ddagger\Ahat\rangle\langle\d\hcal^\ddagger\Bhat\rangle\,,
\eea
where the last term is proportional to $\d t$ and arises from the Wiener noise properties $\d Z_i^*\d Z_j = \delta_{ij}\d t$. This gives the increments for the second Gaussian moments,
\bea\label{eq:du-dv-het}
    \d u_{ij}&= \langle\lcal^\ddagger(\delhat_i\delhat_j)\rangle + \langle\d \hcal^\ddagger(\delhat_i\delhat_j)\rangle\\
    &\phantom{=}- \langle\d\hcal^\ddagger\bbb_i\rangle\langle\d\hcal^\ddagger\bbb_j\rangle\,,\\
        \d v_{ij}&= \langle\lcal^\ddagger(\ddelhat_i\delhat_j)\rangle + \langle\d \hcal^\ddagger(\ddelhat_i\delhat_j)\rangle \\&\phantom{=}- \langle\d\hcal^\ddagger\dbbb_i\rangle\langle\d\hcal^\ddagger\bbb_j\rangle\,.
\eea

\subsection{Quantum-jump dynamics}\label{app:qj-eom}

Similar to our previous treatment, to find the stochastic dynamics governing the Gaussian variational parameters, we proceed by first recasting the quantum-jump equation~\eqref{eq:qj-rho} in its adjoint form,
\bea\label{eq:qj-dO-adjoint}
    \d\langle\Ohat\rangle &= \d t\left\{ \rmi\left\langle[\Hhat,\Ohat]\right\rangle-\dfrac{1}{2} \sum_i\left\langle\hcal^\ddagger[\Lhat_i^\dagger\Lhat_i]\Ohat\right\rangle \right\}\\ &+ \sum_i \d M_i(t) \left\langle \Jcal^\ddagger[\Lhat_i]\Ohat \right\rangle\,,
\eea
where the adjoint superoperators read
\bea
    \hcal^\ddagger[\Ahat]\Ohat&= (\Ahat^\dagger-\langle\Ahat^\dagger\rangle)\Ohat + \Ohat(\Ahat-\langle\Ahat\rangle)\,,\\
\Jcal^\ddagger[\Lhat]\Ohat&=\dfrac{\Lhat_i^\dagger\Ohat\Lhat_i}{\langle\Lhat_i^\dagger\Lhat_i\rangle} - \Ohat\,.
\eea

The equation for $\beta_i$ can be read off directly from the adjoint master equation~\eqref{eq:qj-dO-adjoint}, giving
\bea\label{eq:qj-dbeta-adjoint}
    \d\beta_i &= \d t\left\{ \rmi\left\langle[\Hhat,\bbb_i]\right\rangle-\dfrac{1}{2} \sum_i\left\langle\hcal^\ddagger[\Lhat_i^\dagger\Lhat_i]\bbb_i\right\rangle \right\}\\ &+ \sum_i \d M_i(t) \left\langle \Jcal^\ddagger[\Lhat_i]\bbb_i \right\rangle\,.
\eea
The second moments are again obtained using stochastic differentiation rules. As the adjoint superoperators $\hcal^\ddagger[\dLhat_i\Lhat_i]$ and $\jcal^\ddagger[\Lhat_i]$ share the same properties as $\lcal^\ddagger$ such as discussed in Sec.~\ref{app:het-eom} above, i.e., linearity and blindness to $\mathbb{C}-$numbers, we arrive at,
\bea
    \d\langle\delhat_A\delhat_B\rangle &= \d\langle\Ahat\Bhat\rangle - (\d\langle\Ahat\rangle)\langle\Bhat\rangle - \langle\Ahat\rangle\d\langle\Bhat\rangle - \d\langle\Ahat\rangle\d\langle\Bhat\rangle\\
    &= \d t\left\{ \rmi\left\langle[\Hhat,\delhat_A\delhat_B]\right\rangle-\dfrac{1}{2} \sum_i\left\langle\hcal^\ddagger[\Lhat_i^\dagger\Lhat_i]\delhat_A\delhat_B\right\rangle \right\}\\ &\phantom{==}+ \sum_i \d M_i(t)\left\langle \Jcal^\ddagger[\Lhat_i]\delhat_A\delhat_B \right\rangle\\
    &\phantom{==}-\sum_i\d M_i(t)\langle\jcal^\ddagger[\Lhat_i]\Ahat\rangle\langle\jcal^\ddagger[\Lhat_i]\Bhat\rangle\,,
\eea
where the last term, originating from $\d\langle\Ahat\rangle\d\langle\Bhat\rangle$, is due to the property of the point-process $\d M_i\d M_j=\delta_{ij}\d M_i$. The equations for the covariances then read,
\bea\label{eq:du-dv-qj}
    \d u_{ij} &= \d t\left\{ \rmi\left\langle[\Hhat,\delhat_i\delhat_j]\right\rangle-\dfrac{1}{2} \sum_i\left\langle\hcal^\ddagger[\Lhat_i^\dagger\Lhat_i]\delhat_i\delhat_j\right\rangle \right\}\\ &\phantom{==}+ \sum_i \d M_i(t)\left\langle \Jcal^\ddagger[\Lhat_i]\delhat_i\delhat_j\right\rangle\\
    &\phantom{==}-\sum_i\d M_i(t)\langle\jcal^\ddagger[\Lhat_i]\bbb_i\rangle\langle\jcal^\ddagger[\Lhat_i]\bbb_k\rangle\,,\\
    \d v_{ij} &= \d t\left\{ \rmi\left\langle[\Hhat,\ddelhat_i\delhat_j]\right\rangle-\dfrac{1}{2} \sum_i\left\langle\hcal^\ddagger[\Lhat_i^\dagger\Lhat_i]\ddelhat_i\delhat_j\right\rangle \right\}\\ &\phantom{==}+ \sum_i \d M_i(t)\left\langle \Jcal^\ddagger[\Lhat_i]\ddelhat_i\delhat_j\right\rangle\\
    &\phantom{==}-\sum_i\d M_i(t)\langle\jcal^\ddagger[\Lhat_i]\dbbb_i\rangle\langle\jcal^\ddagger[\Lhat_i]\bbb_j\rangle\,,
\eea
completing the set of equations.

\subsubsection{Implementation with the waiting-time algorithm}
In practice, we follow the standard implementations of the quantum-jump dynamics by sampling the waiting-time distribution between jumps instead of generating $\d M_i$ at every time step~\cite{Wiseman}. To determine the time for each jump event, we first draw a random number $r$ uniformly between 0 and 1, and then integrate the \textit{ordinary} differential equation governing the smooth no-jump evolution,
\bea
    \dfrac{\d\langle\Ohat\rangle}{\d t} &= \rmi\left\langle[\Hhat,\Ohat]\right\rangle-\dfrac{1}{2} \sum_i\left\langle\hcal^\ddagger[\Lhat_i^\dagger\Lhat_i]\Ohat\right\rangle\,,
\eea    
which is equivalent to considering a time evolution generated by the effective non-Hermitian Hamiltonian~\eqref{eq:Heff-nh}. 
This directly translates into an ordinary differential equation for our dynamical variational parameters $Q_i(t)$ [using Eq.~\eqref{eq:dQ}], $u_{ij}(t)$ and $v_{ij}(t)$ using the equations derived above.
The wavefunction norm $\Lambda$, initialized with value $1$,  decays along the non-unitary evolution according to
\bea
    \dfrac{\d \Lambda}{\d t} = -\sum_i\langle\Lhat_i^\dagger\Lhat_i\rangle \Lambda\,,
\eea
and the norm value $\Lambda(t)$ equals the probability of such a no-jump evolution up to time $t$ to occur [see Eq.~\eqref{eq:dm-jump}]. The correct waiting-time distribution is therefore generated by interrupting this evolution at the time where $\Lambda(t)=r$. A jump is then applied, updating the state as
\bea
\langle\Ohat\rangle\to\dfrac{\langle\Lhat_k^\dagger\Ohat\Lhat_k\rangle}{\langle\Lhat_k^\dagger\Lhat_k\rangle}\,,
\eea
where a single jump channel $k$ is selected with probability proportional to $\langle\Lhat_k^\dagger\Lhat_k\rangle$. In terms of the Gaussian parameters of the state ansatz, this update reads [see Eq.~\eqref{eq:du-dv-qj}],
\bea\label{eq:jump-update-naive}
    \beta_i & \to \beta_i^J\equiv \dfrac{\langle\Lhat^\dagger_k\bbb_i\Lhat_k\rangle}{\langle\dLhat_k\Lhat_k\rangle}\,,\\
    u_{ij} & \to \dfrac{\langle\dLhat_k\bbb_i\bbb_j\Lhat_k\rangle}{\langle\dLhat_k\Lhat_k\rangle} - \beta_i^J\beta_j^J\,,\\
    v_{ij} & \to \dfrac{\langle\dLhat_k\dbbb_i\bbb_j\Lhat_k\rangle}{\langle\dLhat_k\Lhat_k\rangle} - \beta_i^{J*}\beta_j^J\,,\\
    Q_i & \to R(\beta^J_i)Q_i\,,
\eea
where the last equation for the quaternion $Q_i$ uses the finite (non-infinitesimal) version of the rotation defined in Eq.~\eqref{eq:Rbeta}, which updates the local comoving frames and resets the dummy variables $\beta_i$ to 0. This completes the jump update, after which we reset the norm to $\Lambda=1$ and start over with a new iteration of the above waiting-time sampling procedure, until the desired final time of the evolution is reached. 

\subsubsection{Implementation of local jumps}\label{app:local-qj}
The above formulation of quantum-jump spin-wave dynamics, despite being formally generic, yields only a rough approximation if implemented naively. For example, the jump operator $\Lhat_i=\sqrt{\gamma}\sigmahat^-_i$, which sends any local spin state to the down state, inevitably generates a large displacement $\beta_i^J$ in the bosonized dynamics (and hence an uncontrolled spin-wave density) when a jump occurs, eventually breaking the validity of the spin-wave approximation. Luckily, since the post-jump state (for this specific jump process) of the site corresponding to the jump channel is known, namely the $\ket\downarrow_z$ state disentangled from the rest of the system, one can encode this information directly in the jump update. This results in a correction to the update rules~\eqref{eq:jump-update-naive}: assuming $k$ is the selected jump channel, we set all covariances involving site $k$, i.e. those of the form $u_{ki}$, $u_{ik}$, $v_{ki}$, $v_{ik}$, to zero, and update $Q_k$ to a quaternion corresponding to the down-pointing local frame, for example $Q_k\to(0,1,0,0)$. In weakly entangled systems, this locally exact correction significantly improves the validity of the spin-wave approximations.  

\newpage
\bibliography{BiblioEPT_Semi}
 
\end{document}